\documentclass [a4paper,10pt,english]{article} 
\usepackage[top=1.2in, bottom=0.8in, left=.6in, right=.6in]{geometry}
\usepackage{amsmath} 
\usepackage[utf8]{inputenc} 
\usepackage[bookmarks=false, colorlinks=false,unicode]{hyperref}
\usepackage[T1]{fontenc}
\usepackage{hyperref} 
\usepackage{graphicx} 
\usepackage{listings} 
\usepackage{subfigure}                  
\usepackage{physics}
\usepackage{amsmath, amsthm, amssymb, amsfonts}
\usepackage{authblk}
\usepackage{blindtext}
\usepackage{xcolor}

\newcommand{\rr}{\mbox{\boldmath $r$}}

\newcommand{\rb}{\mbox{\boldmath $b$}}

\newcommand{\rd}{\mbox{\boldmath $\Delta$}}

\begin{document}

\title{Coherent and incoherent vector meson electroproduction in the future electron - ion colliders: the hot - spot predictions}
\author[1,2]{M. Krelina}
\author[3]{V.P. Goncalves}
\author[2]{J. Cepila}

\affil[1]{
Departamento de F\'{\i}sica,
Universidad T\'ecnica Federico Santa Mar\'{\i}a,
Casilla 110-V, Valpara\'{\i}so, Chile
}
\affil[2]{
Czech Technical University in Prague, FNSPE, B\v rehov\'a 7, 11519
Prague, Czech Republic
}
\affil[3]{
Instituto de F\'{\i}sica e Matem\'atica, Universidade Federal de Pelotas,
Caixa Postal 354, CEP 96010-900, Pelotas, RS, Brazil
}

\maketitle

\begin{abstract}
One of the more promising observables to  probe the high energy regime of the QCD dynamics in the future Electron - Ion  Colliders (EIC) is the exclusive vector meson production cross section in coherent and incoherent interactions. Such processes measure the average spatial distribution of gluons in the target as well the fluctuations and correlations in the gluon density. In this paper we present a comprehensive analysis of the energy, photon virtuality, atomic number and momentum transfer dependencies of the coherent and incoherent cross sections considering two different models for the nuclear profile function. In particular, we present the predictions of the hot - spot model, which assumes the presence of subnucleonic degrees of freedom and an energy-dependent profile. Our results indicate that the analysis of the ratio between the incoherent and coherent cross sections and the momentum transfer distributions in the future EIC can be useful to constrain the description of the hadronic structure at high energies.

\end{abstract}

%
%

\section{Introduction}

One of the main goals of the future Electron - Ion Colliders (EIC) \cite{Raju,Boer,Accardi,LHeC,Aschenauer:2017jsk} is to perform a detailed investigation of the hadronic structure in the non-linear regime of the Quantum Chromodynamics (QCD) and, in particular, to be able to determine the presence of gluon saturation effects, the   magnitude of the associated non-linear corrections and what is the correct theoretical framework for their description \cite{hdqcd}. Such expectations are motivated by the enhancement of the non-linear effects with the nuclear mass number through the nuclear saturation scale, $Q^2_{s,A}$, which determines the onset of non-linear effects in the QCD dynamics, being  enhanced with respect to the nucleon one by a factor $ \propto A^{\frac{1}{3}}$. Consequently, it is possible to access
in electron - nucleus ($eA$) collisions the high parton densities that would be achieved  in an electron - proton collider at energies that are at least one order of magnitude higher those probed at HERA.

\begin{figure}%
\centering
\subfigure{%
\includegraphics[width=0.48\textwidth]{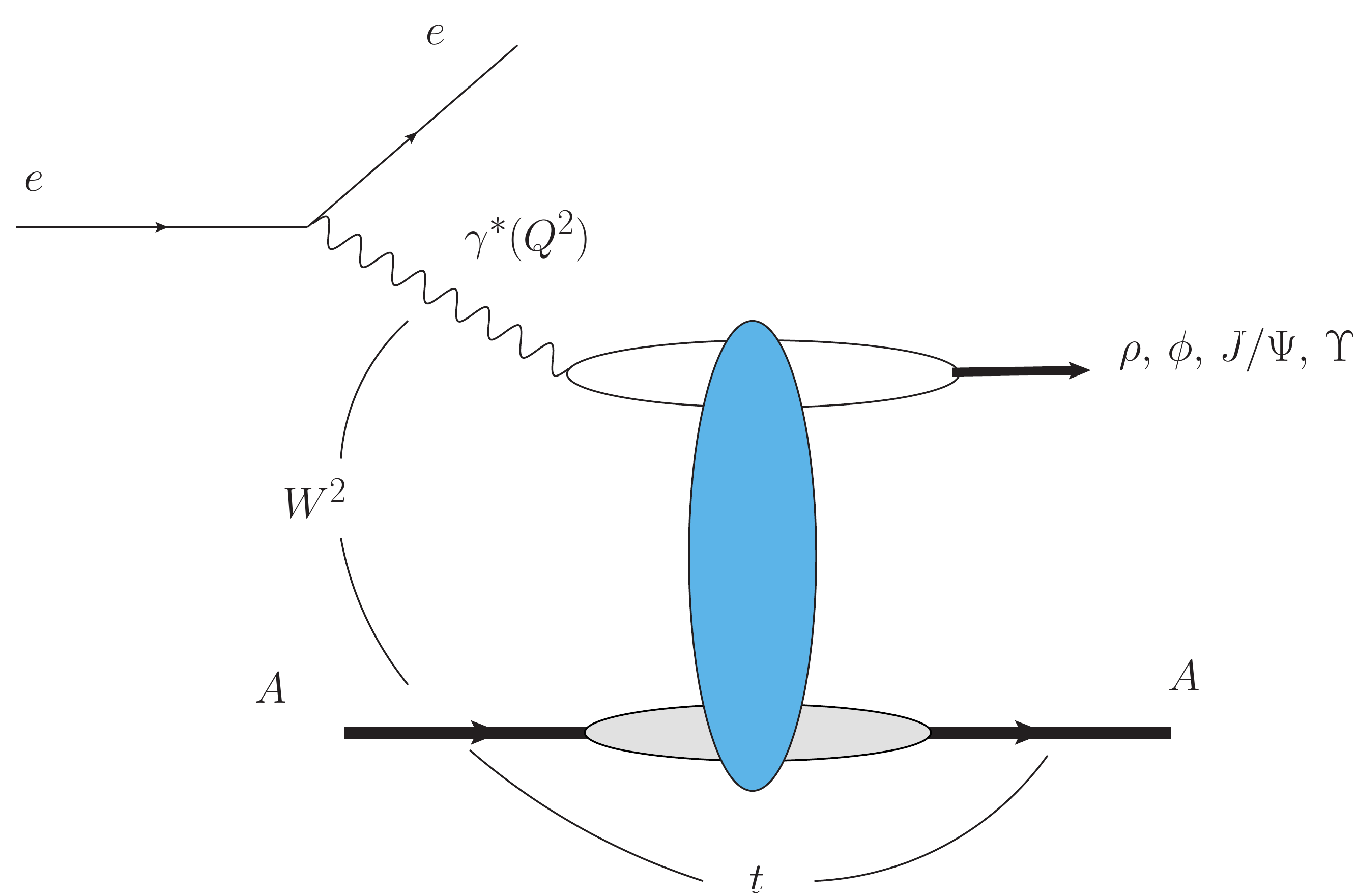}}%
\qquad
\subfigure{%
\includegraphics[width=0.48\textwidth]{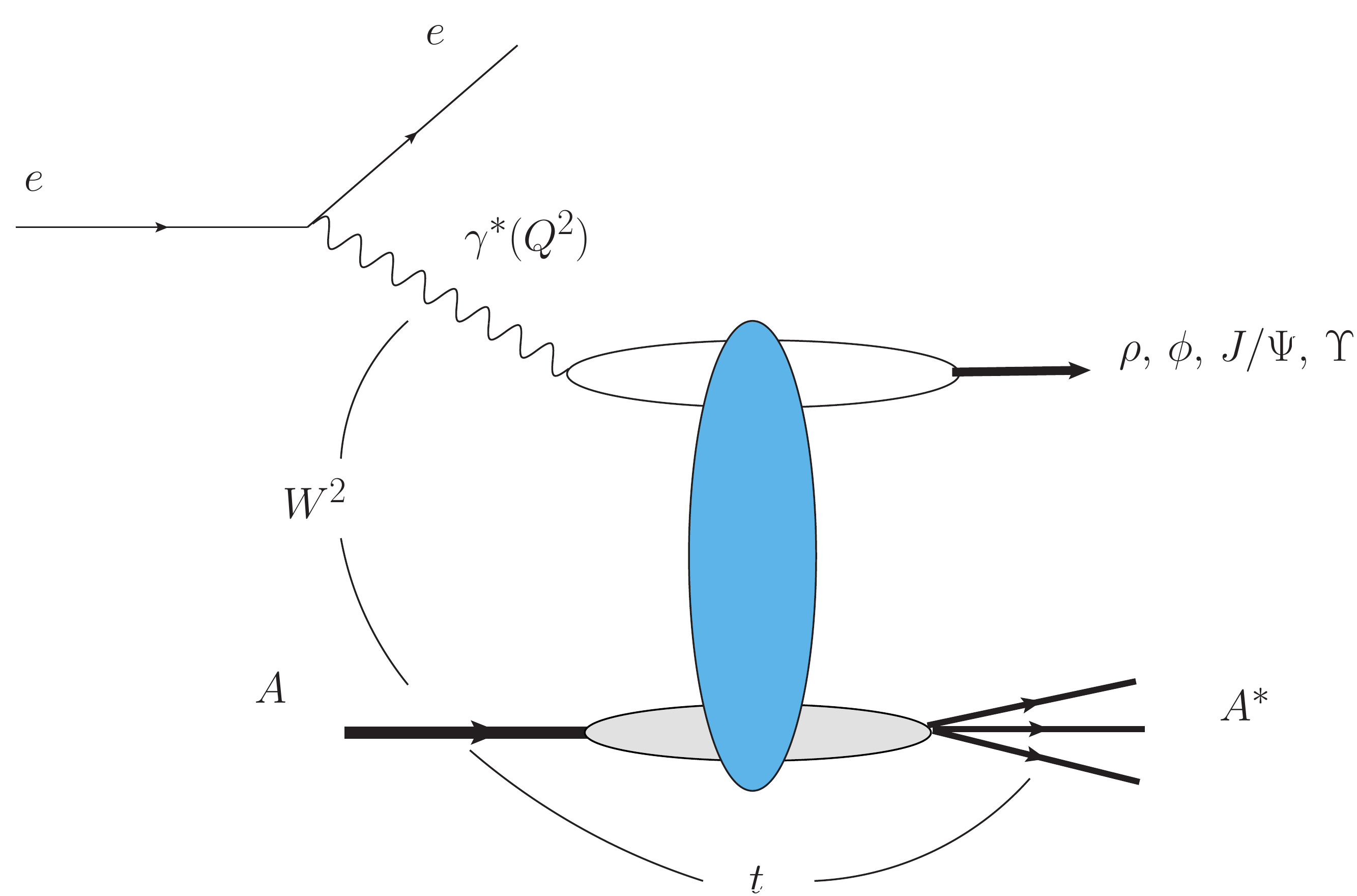}}
\label{fig:diagram}%
\caption{The coherent (left) and incoherent (right) exclusive vector meson production in $eA$ collisions.}
\end{figure}

A smoking gun of the gluon saturation effects in $eA$ collisions is the analysis of
diffractive events, which are predicted to contribute with half of the total cross section in the asymptotic limit of very high energies, with the other half being formed by all inelastic processes \cite{Nikolaev,simone2}. For the kinematical range of the future EIC, it is expected that the contribution of the diffractive events is $\approx 20 \%$ \cite{Nik_schafer,Kowalski_prc,erike_ea2}, which have motivated
an intense phenomenology about the implications of the gluon saturation effects in the diffractive production of different final states.
A promising observable is the exclusive vector meson production off large nuclei
\cite{vmprc,Caldwell,Lappi_inc,Toll,Lappi_bal,diego,Mantysaari:2017slo,Mantysaari:2018zdd}.
In the QCD dipole approach \cite{nik}, such process can be factorized in terms of the fluctuation of the virtual photon into a $q \bar{q}$ color
dipole, the dipole-nucleus scattering by a color singlet exchange and the recombination into the exclusive final state. An important characteristic of the exclusive vector meson production is that it is experimentally clean, with the final state being unambiguously identified by the presence of a rapidity gap. Moreover, such processes are driven by the gluon content of the target. As the cross section is proportional
to the square of the scattering amplitude,  the exclusive vector meson production is strongly sensitive to the underlying QCD dynamics. Another advantage of the study of this process in $eA$ collisions is the possibility of the study of coherent and incoherent interactions, which provide different insights about the nuclear structure and the QCD dynamics at high energies. The coherent and incoherent vector meson production in $eA$ collisions are represented in Fig. \ref{fig:diagram}.  If the nucleus scatters elastically,  the process is called coherent production, and the associated cross section measures the average spatial distribution of gluons in the target. On the other hand,  if the nucleus scatters inelastically, i.e., breaks up due to the $p_T$ kick given to the nucleus, the process is denoted incoherent production.
 In this case, one sums over all final states of the target nucleus,
except those that contain particle production. The associated cross section probes the fluctuations and correlations in the gluon density.
In both cases, the final state is characterized by a rapidity gap.
It is expected that the coherent production dominates at small squared transverse momentum transfer $t$ ($|t|\cdot R_A^2/3 \ll 1$, where $R_A$ is the nuclear radius), with its signature being a sharp
forward diffraction peak. On the other hand, incoherent production should dominate at large $t$ ($|t|\cdot R_A^2/3 \gg 1$), with the associated $t$-dependence being to a good accuracy the same as in the production off free nucleons. As the momentum transfer is Fourier conjugate to the impact parameter, the coherent and incoherent exclusive vector meson production are sensitive to different aspects of the geometric structure of the target, which at high energies can be identified with the spatial gluon distribution of the target. In the coherent case, the averaged density profile of the gluon density is probed. In contrast, the incoherent cross sections constrain the event - by - event fluctuations of the gluonic fields in the target.

Our goal in this paper is to present a detailed investigation of the coherent and incoherent exclusive vector meson electroproduction in $eA$ collisions considering the energy-dependent hot -- spot model proposed in Ref. \cite{Cepila:2016uku} for a proton target and extended for the nuclear case in Refs. \cite{Cepila:2017nef,Cepila:2018zky} (For similar approaches see, e.g. Refs.
\cite{Mantysaari:2016ykx,Mantysaari:2016jaz,Traini:2018hxd}). In this model, the hadronic structure is described in terms of subnucleonic degrees of freedom representing regions of high gluon density, denoted hot -- spots, which increase in number with the decreasing of the Bjorken - $x$ variable.  Such energy dependence is motivated by the fact that the non - linear QCD dynamics predicts that the transverse density profile of the target change with the energy. As demonstrated in Refs. \cite{Cepila:2017nef,Cepila:2018zky,Bendova:2018bbb}, such model is able to describe the current data for the exclusive and dissociative production of vector mesons in $ep$ collisions, as well find a satisfactory agreement with the data for the exclusive $J/\Psi$ photoproduction in ultraperipheral heavy ion collisions. In this paper we will estimate the coherent and incoherent cross sections for the production of light ($\rho$ and $\phi$) and heavy ($J/\Psi$ and $\Upsilon$) vector mesons considering different nuclear targets ($A = Au, Xe$ and $Ca$) and assuming two distinct models for the nuclear profile.  We will present predictions for the dependencies of the cross sections on the energy, atomic number, photon virtuality and squared momentum transfer. Our results demonstrate that the ratio between the incoherent and coherent cross sections is strongly sensitive to the presence of subnucleonic degrees of freedom in the form of hot spots.

This paper is organized as follows. In the next Section, we present a brief review of the formalism and discuss the two models for the nuclear profile used in our calculations. In Section \ref{sec:results} we present our results for the coherent and incoherent cross sections, considering the kinematical range that will be probed by the electron -- ion facilities that are under design: the EIC in the USA and the LHeC project at CERN. Finally, in Section \ref{sec:sum} we summarize our main conclusions.

%
%

\section{Review of the formalism}
The coherent and incoherent exclusive vector meson electroproduction in $eA$ collisions are represented in the left and right panels of the Fig. \ref{fig:diagram}, respectively. The reaction is given by $e(l) + A(P) \rightarrow e(l^\prime) + Y(P^\prime) + V (P_V)$, where $Y = A$ in the coherent case and $Y = A^*$ for incoherent interactions. Moreover, $l$ and $l^\prime$ are the electron momenta in the initial and final state, respectively, while $P$ and $P^\prime$ are the inital and final nucleus momenta. Finally, $P_V$ is the momentum of the vector meson in the final state. The kinematics is described by the following Lorentz invariant quantities: $Q^2 = - q^2 = - (l - l^\prime)^2$, $t = - (P^\prime - P)^2$, $W^2 = (P + q)^2$ and $x = (M^2 + Q^2 -t)/(W^2+Q^2)$ (note, this definition of $x$ differs from the one in \cite{Cepila:2016uku,Cepila:2017nef}), where $Q^2$ is the photon virtuality, $W$ is the center of mass energy of the virtual photon -- nucleus system and $M$ is the mass of the vector meson.
In the color dipole formalism,  the  $eA \rightarrow e V Y$ process can be factorized in terms of the fluctuation of the virtual photon into a $q \bar{q}$ color dipole, the dipole-nucleus scattering by a color singlet exchange  and the recombination into the exclusive final state $V$. The  amplitude for producing a vector meson diffractively in an electron - nucleus scattering is given by
\begin{eqnarray}
 {\cal A}_{T,L}({x},Q^2,\Delta)  =  i
\,\int d^2\rr \int \frac{dz}{4\pi} \int \, d^2\rb \, e^{-i[\rb -(1-z)\rr].\rd}
 \,\, (\Psi^{V*}\Psi)_{T,L}  \,\,\frac{d\sigma_{dA}}{d^2\rb}({x},\rr,\rb)
\label{amp}
\end{eqnarray}
where $T$ and $L$ denotes the transverse and longitudinal polarizations of the virtual photon, $(\Psi^{V*}\Psi)_{i}$ denotes the wave function overlap between the virtual photon and vector meson wave functions, $\Delta = \sqrt{-t}$ is the momentum transfer and $\rb$ is the impact parameter of the dipole relative to the target. The variables  $\rr$ and $z$ are the dipole transverse radius and the momentum fraction of the photon carried by a quark (an antiquark carries then $1-z$), respectively. Moreover,  ${d\sigma_{dA}}/{d^2\rb}$  is the dipole-nucleus cross section (for a dipole at  impact parameter $\rb$) which encodes all the information about the hadronic scattering, and thus about the non-linear and quantum effects in the hadron wave function.
Such quantity  depends on the $\gamma^*A$ center - of - mass reaction energy, $W$, the photon virtuality and mass of the vector meson, through the Bjorken - $x$ variable.  Consequently,  the cross section for the production of light vector mesons  at low $Q^2$ is much more sensitive to low $x$ effects than the  one for  production of heavy mesons.  In addition, the study of the $\rho$ and $\phi$ production at different photon virtualities allows to probe the transition between the non-linear and linear regimes of the QCD dynamics. In principle, ${d\sigma_{dA}}/{d^2\rb}$ can be derived using the
Color Glass Condensate (CGC) formalism \cite{CGC}, which is characterized by the infinite hierarchy of equations, the so called Balitsky-JIMWLK equations \cite{BAL,CGC}, which reduces in the mean field approximation to the Balitsky-Kovchegov (BK) equation \cite{BAL,kov}.

As in analysis presented in Ref. \cite{Cepila:2017nef}, in this paper we will describe the dipole - nucleus cross section derived using the
Glauber-Gribov formalism \cite{gribov}, which is given by
\begin{eqnarray}
\frac{d\sigma_{dA}}{d^2\rb} = 2\,\left( 1 - \exp \left[-\frac{1}{2}  \, \sigma_{dp}(x,\rr^2) \,T_A(\rb)\right]\right) \,\,,
\label{enenuc}
\end{eqnarray}
where $\sigma_{dp}$ is the dipole-proton cross section and $T_A(\rb)$ is  the nuclear profile function. Such equation takes into  account the
multiple elastic rescattering diagrams of the $q \overline{q}$ pair and  is justified  in the large coherence length regime ($l_c \gg R_A$). In this limit the transverse separation $\rr$ of partons in the multiparton Fock state of the photon becomes a conserved quantity, {\it i.e.},  the size of the pair $\rr$ becomes eigenvalue
of the scattering matrix. Following Refs. \cite{Cepila:2016uku,Cepila:2017nef}, we will assume that $\sigma_{dp}(x,\rr^2) = \sigma_0 \, {\cal{N}}_p(x,\rr^2)$,
where the value of $\sigma_0$ is fixed by the value of the proton profile in impact parameter space and ${\cal{N}}_p$ is forward dipole scattering amplitude, which we chose to be given by the model of Golec-Biernat and
Wusthoff \cite{GBW,Cepila:2016uku,Cepila:2018zky}
\begin{equation}\label{eq:N}
  {\cal{N}}_p (x,\rr^2)  = (1 - \exp[ -r^2 Q^2_s(x)/4]), \quad Q^2_s(x) = Q^2_0(x_0/x)^\lambda
\end{equation}
with the saturation scale, $Q^2_s$, given by the parameters $\lambda, x_0$ and $Q_0^2$.
 It is important to emphasize that this model for ${d\sigma_{dA}}/{d^2\rb}$ allows to describe the current  experimental data on the nuclear
structure function \cite{armesto,Kowalski_prc,erike_ea2}.
In order to estimate the amplitude ${\cal A}_{T,L}({x},Q^2,\Delta)$ we also should to assume a model for the overlap function $(\Psi^{V*}\Psi)_{i}$.
In what follows we will assume the boosted-Gaussian model for the vector meson wave functions \cite{Nemchik:1994fp,Nemchik:1996cw}, with the numerical values of
the parameters as in \cite{Kowalski:2006hc}.

For coherent interactions, the nucleus is required to remain in its ground state, i.e., intact after the interaction, which corresponds to take the average over the configurations of the nuclear wave function at the level of the scattering amplitude. Consequently, the coherent cross section is obtained by averaging the amplitude before squaring it and the differential distribution will be given by
\begin{equation}\label{eq:xsec-coh}
  \left.\frac{d\sigma^{\gamma A \rightarrow V\,A}}{dt}\right|_{T,L} =
  \frac{1}{16\pi}\left| \left\langle \mathcal{A}(x,Q^2, \Delta)_{T,L} \right\rangle \right|^2.
\end{equation}
On the other hand, for incoherent interactions the average over configurations is at the cross section level, the nucleus can break up and the resulting incoherent cross section will be proportional to the variance of the amplitude with respect to the nucleon configurations of the nucleus, i.e., it will measure the fluctuations of the gluon density inside the nucleus. In this case, the differential cross sections will be expressed as follows:
\begin{equation}\label{eq:xsec-inc}
  \left.\frac{d\sigma^{\gamma A \rightarrow V\,Y}}{dt}\right|_{T,L} = \frac{1}{16\pi}
  \left(  \left\langle\left|  \mathcal{A}(x,Q^2,\vec \Delta)_{T,L}  \right|^2 \right\rangle  - \left| \left\langle \mathcal{A}(x,Q^2,\vec \Delta)_{T,L} \right\rangle \right|^2\right),
\end{equation}
where $Y = A^*$ represents the dissociative state. In our calculations we will include the skewedness correction by multiplicating the coherent and incoherent cross sections by the factor $(R_g^{T,L})^2$ as given in Ref. \cite{Shuvaev:1999ce}.

\begin{figure}[!tbp]
  \centering
  \begin{minipage}[b]{0.325\textwidth}
    \includegraphics[width=\textwidth]{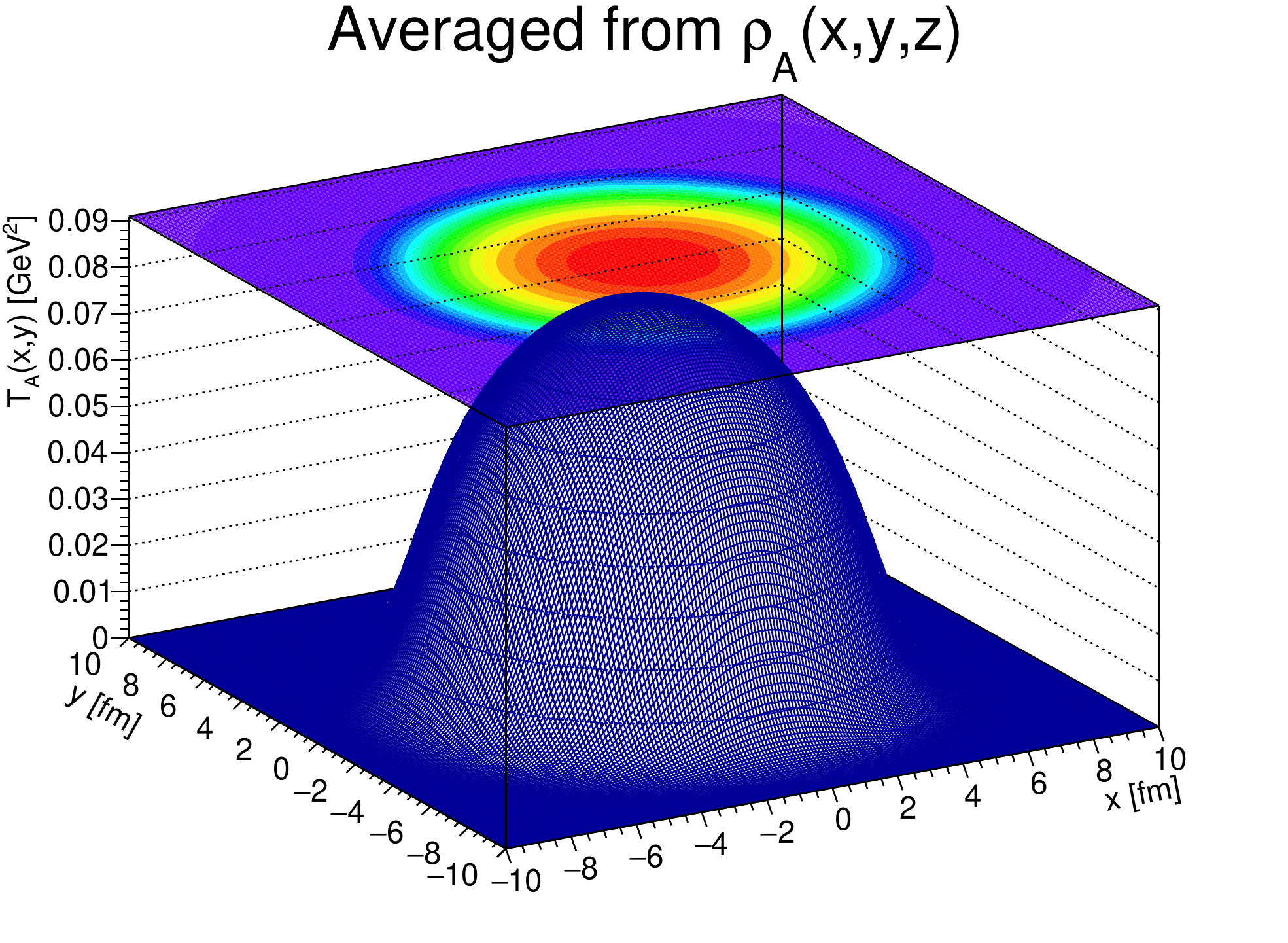}
  \end{minipage}
  \hfill
  \begin{minipage}[b]{0.325\textwidth}
    \includegraphics[width=\textwidth]{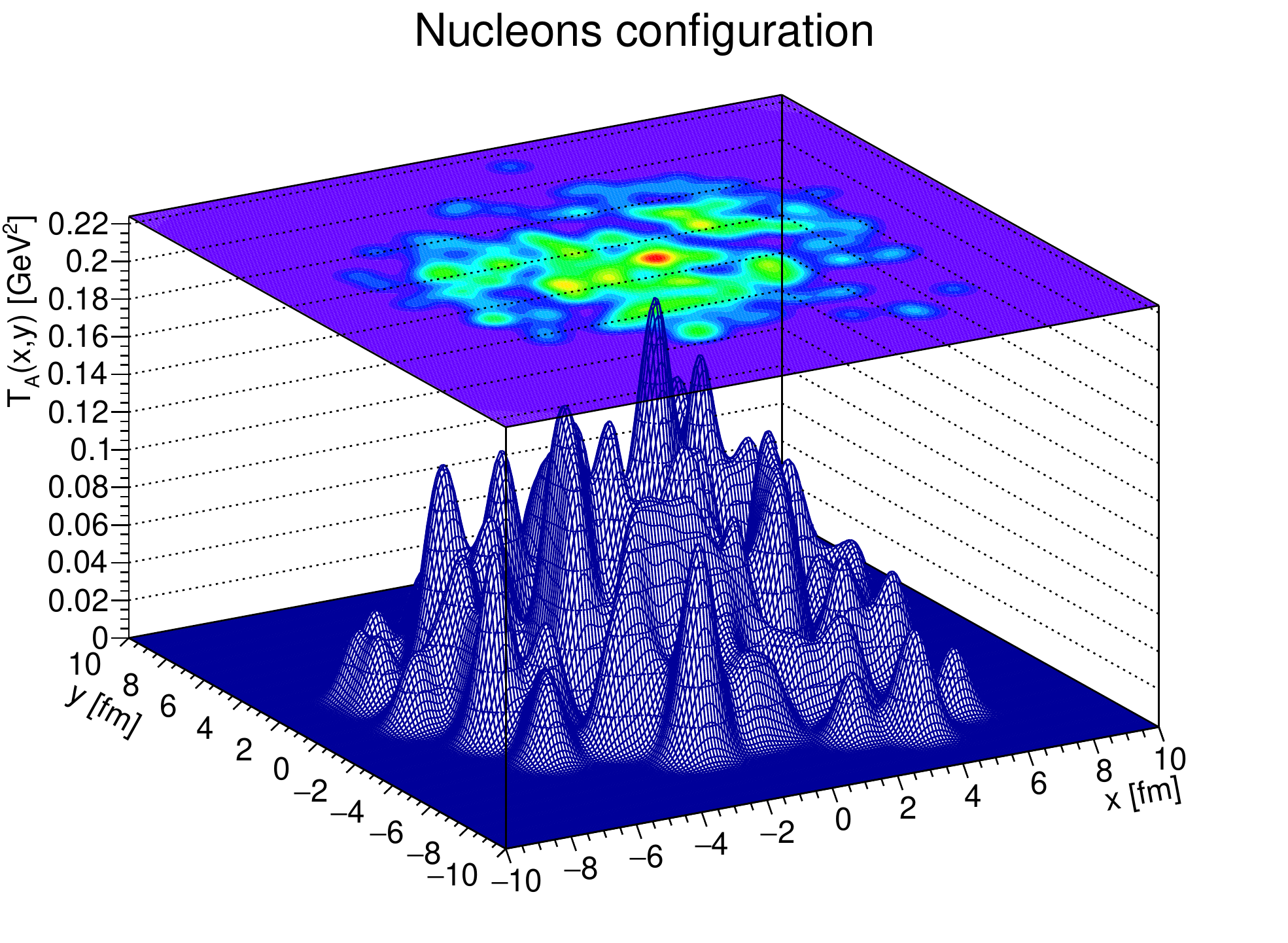}
  \end{minipage}
  \hfill
  \begin{minipage}[b]{0.325\textwidth}
    \includegraphics[width=\textwidth]{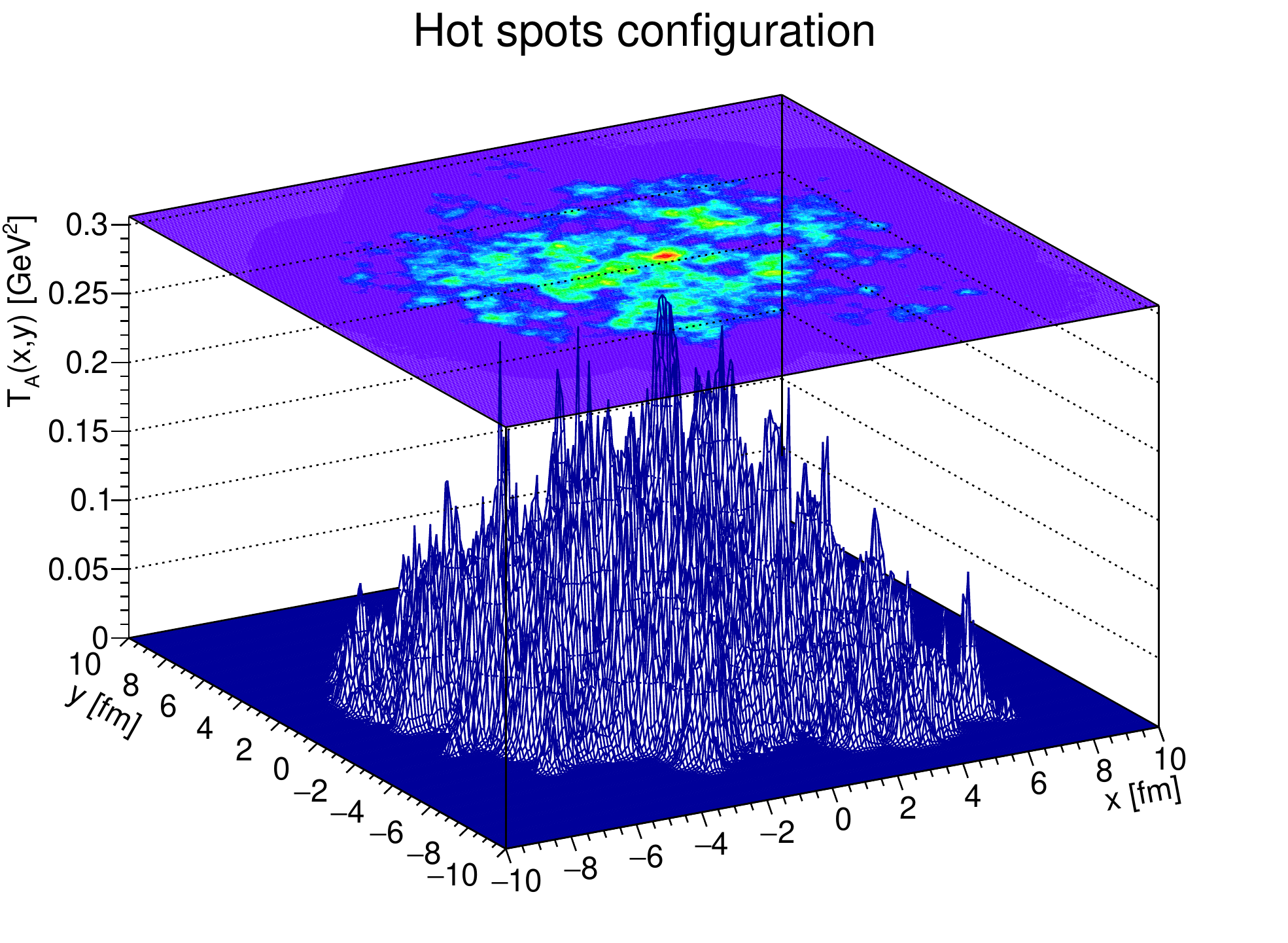}
  \end{minipage}
  \caption{Comparison of nuclear profiles, $T_{A}(b)$, calculated from nuclear density function $\rho_A(\vec r)$ (left), calculated using $nu$ model where each nucleon is represented by Gaussian (center), and calculated using the hot shots model, where every nucleon from $nu$ model is represented by a set of hot spots (right).}
  \label{fig:TAcompareMap}
\end{figure}

The coherent and incoherent cross sections depend on the description of the nuclear profile $T_A(\rb)$. It is useful in the literature to estimate this quantity assuming a given model for the nuclear density function $\rho_A (\vec{r})$, which implies the smooth behaviour for $T_A(\rb)$ represented in the left panel of Fig. \ref{fig:TAcompareMap}. In principle, such model is realistic if the observable  that we are interested is sensitive only to the averaged behaviour over the configurations in the nuclear wave function. However, as discussed above, the incoherent cross section is sensitive to fluctuations in the configurations. Therefore, a more detailed model should be considered to estimate this observable. One possibility is to assume that each nucleon in the nucleus  has a Gaussian profile of width $B_p$, centered at random positions $\rb_i$ sampled from a Woods-Saxon nuclear profile
\begin{equation}\label{eq:Ths0}
  T_A(\rb) = \frac{1}{2\pi B_p} \sum_{i=1}^{A} \exp\left[ - \frac{(\rb - \rb_i)^2}{2B_p} \right] \,\,.
\end{equation}
A typical configuration for this model, denoted {\it nu} model hereafter, is represented in the central panel of Fig.  \ref{fig:TAcompareMap}.
On the other hand, we also can assume that the nucleons inside the nucleus are themselves made up of hot spots with a Gaussian profile of width $B_{hs}$, distributed according to a Gaussian of width $B_{p}$ inside the nucleon at arbitrary position $\rb_j$
\begin{equation}\label{eq:Ths}
  T_A(\rb) = \frac{1}{2\pi B_{hs}} \sum_{i=1}^{A} \frac{1}{N_{hs}} \sum_{j=1}^{N_{hs}} \exp\left[ - \frac{(\rb - \rb_i - \rb_j)^2}{2B_{hs}} \right]
\end{equation}
where in this case $N_{hs}$ is a random number drawn from a zero-truncated Poisson distribution \cite{Cepila:2017nef},
where the Poisson distribution has a mean value
\begin{equation}\label{eq:Nhs}
  \langle N_{hs}(x) \rangle = p_0 x^{p_1} (1 + p_2 \sqrt{x})\,\,.
\end{equation}
A typical configuration of the hot - spot ({\it hs}) model  is presented in the right panel of Fig.
 \ref{fig:TAcompareMap}.
Some additional comments are in order. In the left panel, nuclear profile is calculated from the nuclear density $\rho_A(\vec r)$ function (for Gold) by integration over longitudinal coordinate $z$. Since it is an average, this nuclear profile has a zero variance for event-by-event calculation. The nuclear profile of the $nu$ model (central panel) is represented by $A$ Gaussians seeded according to the distribution $\rho_A(\vec r)$. This model leads to different configuration for every event, as was suggested, e.g. in \cite{Kowalski:2006hc, Mantysaari:2017dwh}, leading to the non-zero variance. Finally, in the $hs$ model (right panel),  for every nucleon of the $nu$ model are generated hot spots whose number is controlled by Eq. \eqref{eq:Nhs} (here for $x=0.001$).  If we compare both visualizations  we can find that the  $hs$ model predicts a more dilute and non - uniform distribution in comparison to the $nu$ one.

In what follows we will estimate the coherent and incoherent cross sections considering the {\it nu} and {\it hs} models of the profile nuclear function. Our goal is to verify if these cross sections are able to discriminate between these two models for the description of the nucleon configurations in the nucleus and if they are sensitive to the presence of hot - spots inside the nucleons. All parameters present in our calculations  have been fixed from a comparison to data on $J/\psi$ and $\rho$
photoproduction off protons \cite{Cepila:2018zky}. The values of the parameters and the associated discussion can be found in \cite{Cepila:2016uku} and will not be repeated here.
Particularly, we would like to highlight the discussion of the chosen values of $B_p$ according to the studied vector mesons in \cite{Cepila:2017nef,Cepila:2018zky,Bendova:2018bbb}.

 Some comments are in order. In our study, following the previous studies performed in Refs.
\cite{Cepila:2016uku,Cepila:2017nef,Cepila:2018zky} we are assuming that $B_p$ and $B_{hs}$ are constants, with its values being determined by fitting the HERA data. As demonstrated in \cite{Cepila:2016uku,Cepila:2017nef,Cepila:2018zky}, such simplified approach describes the experimental data for the total cross sections and $t$-distributions. In principle, such quantities can be energy dependent. An alternative is to assume that $B_{hs}$ is related to the saturation scale by $B_{hs} = 1/Q_s^2(x)$, which implies that at larger energies we will have smaller hot-spots. Moreover, it is also possible to assume that the radius of the nucleons has a  logarithmic growth with energy. This energy dependence will imply  a
change of the number of hot-spots, such that when the nucleon is larger,
more hot spots will be needed to fill the phase space. In other words, there is a correlation between $B_p$ and $N_{hs}$. The study of such alternatives are the subject of a separate publication. The impact on our predictions for large nuclear targets is small.
Finally, a comment about the numerical method used in our calculations is important. The mean value in Eqs. \eqref{eq:xsec-coh} and \ref{eq:xsec-inc} refers to the average and the variance over the configurations (events). Due to the enormous computer resources needed to evaluate the integrals in Eq. \eqref{amp}, we have only used 200 configurations. The error from the integration of individual configuration is low ($\approx 2\%$). However, the statistical error of the set of 200 configurations is higher roughly 10\% for incoherent cross section (coherent cross section error is approx. 3\% ), but for $\rho$ meson with increasing nucleon number $A$ and decreasing scale $Q^2$ can reach up to 30\% for Gold and $Q^2=0.05$~GeV$^2$.  In our predictions we will present the associated uncertainty bands.

%
%

\section{Results}
\label{sec:results}

In what follows we will present our predictions for the energy, photon virtuality, atomic number and momentum transfer of the coherent and incoherent cross sections considering the two models for the nuclear profile function discussed in the previous Section. We will consider the production of two light vector mesons ($\rho$ and $\phi$) as well two heavy vector mesons ($J/\Psi$ and $\Upsilon$). Our focus will be in the kinematical range that probably will probed in the future electron - ion colliders under design: the EIC in USA ($\sqrt{s} \approx 100$ GeV) and the LHeC at CERN ($\sqrt{s} \approx 1000$ GeV). As we are interested in the low - $x$ region, we will present results for $Q^2 \le 10$ GeV$^2$ and will consider different atomic nuclei ($A = Au, \, Xe$ and $Ca$).

\begin{figure}%
\centering
\subfigure{%
\includegraphics[width=0.4\textwidth]{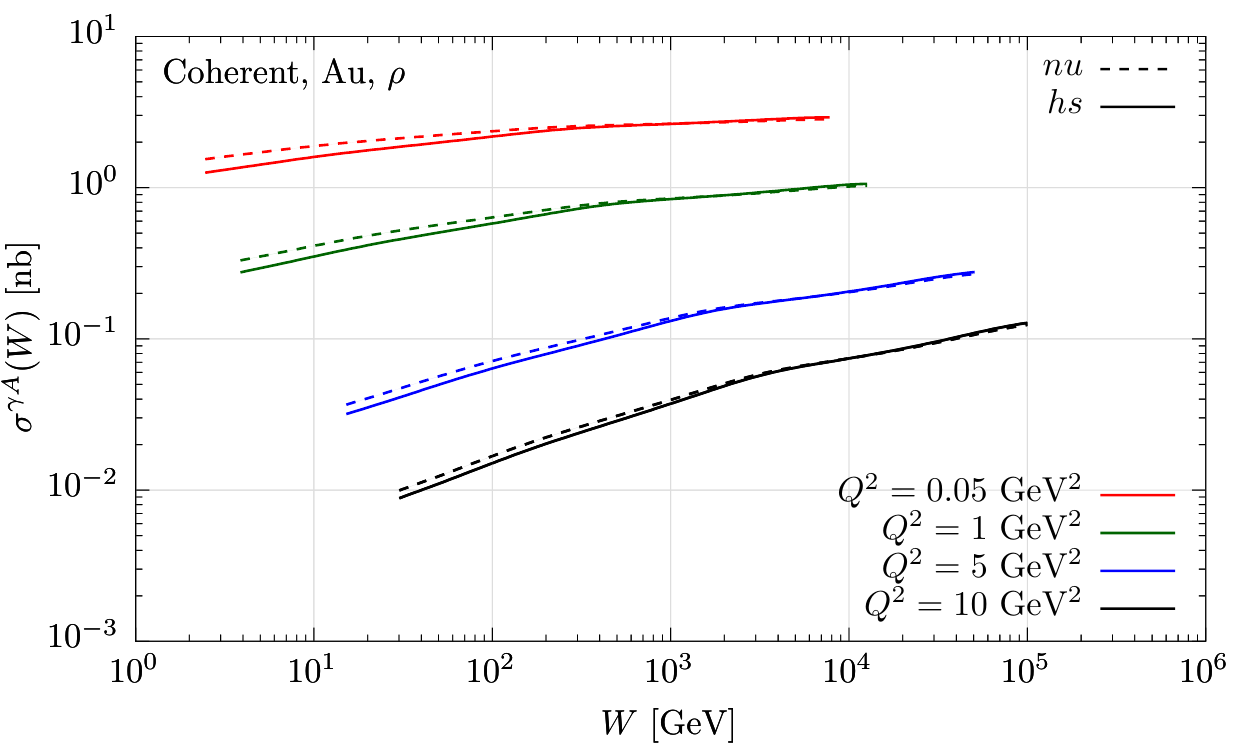}}%
\qquad
\subfigure{%
\includegraphics[width=0.4\textwidth]{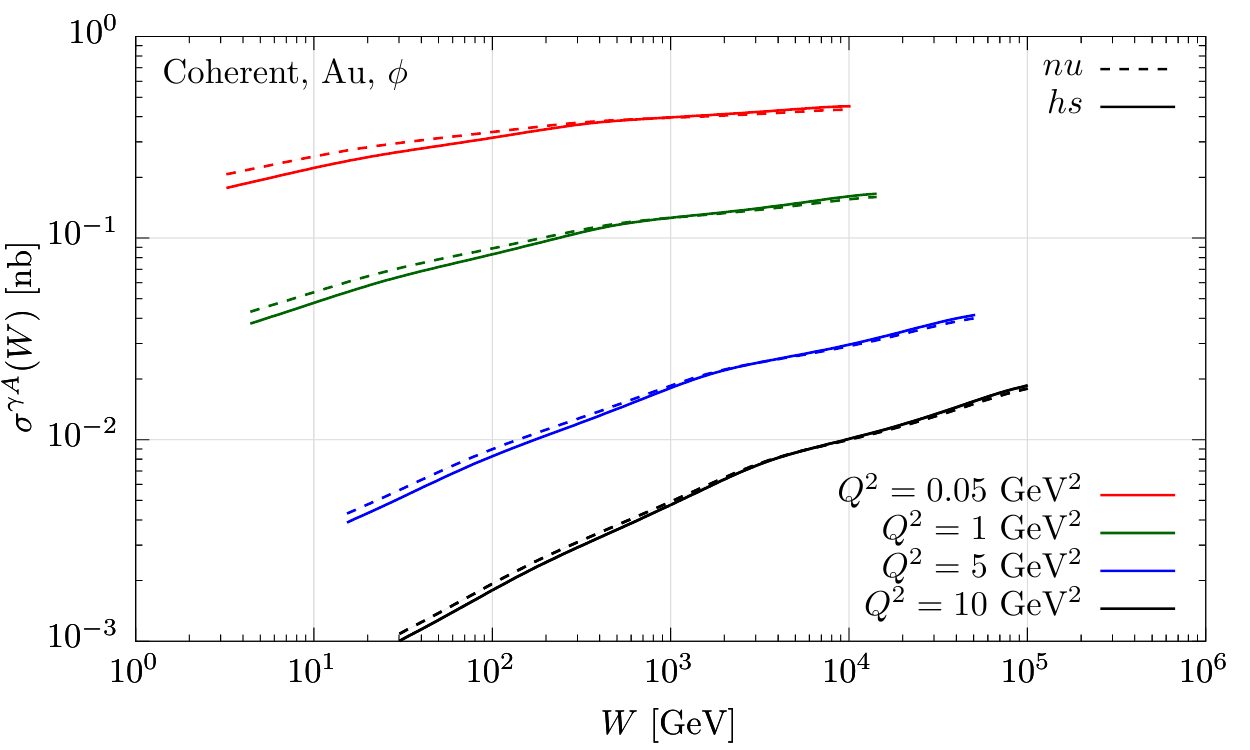}}\\%
\subfigure{%
\includegraphics[width=0.4\textwidth]{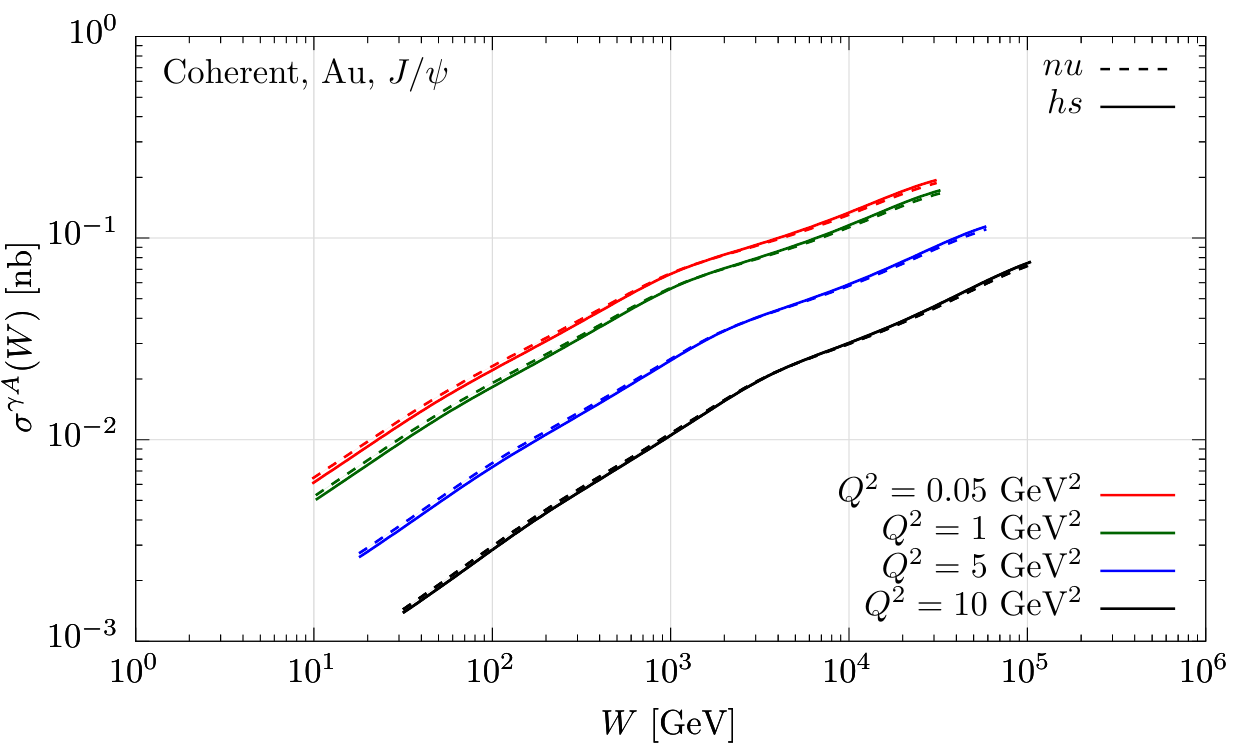}}%
\qquad
\subfigure{%
\includegraphics[width=0.4\textwidth]{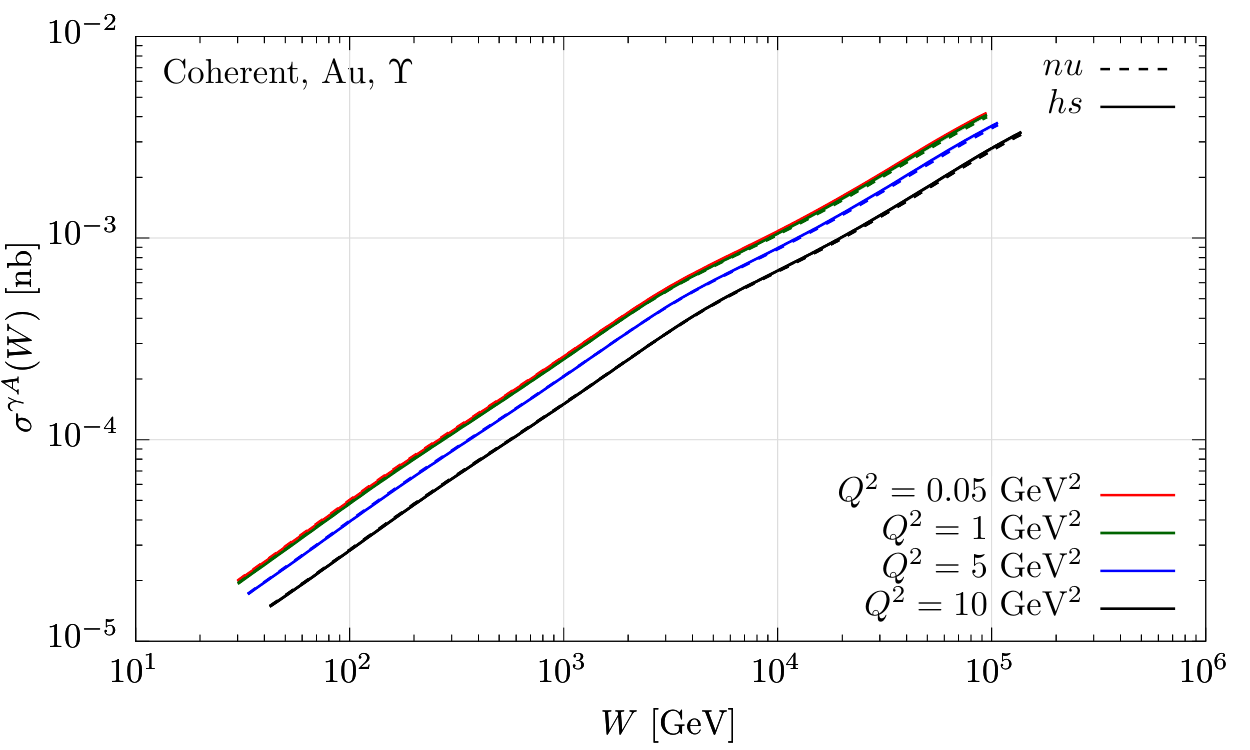}}%
\caption{Predictions for coherent exclusive vector meson production considering a Gold target as a function of energy $W$ for different values of $Q^2$. The solid (dashed) lines correspond to the predictions of the $hs$ ($nu$) model for the nuclear profile.}
\label{fig:xseccoh-energy}%
\end{figure}

\begin{figure}%
\centering
\subfigure{%
\includegraphics[width=0.4\textwidth]{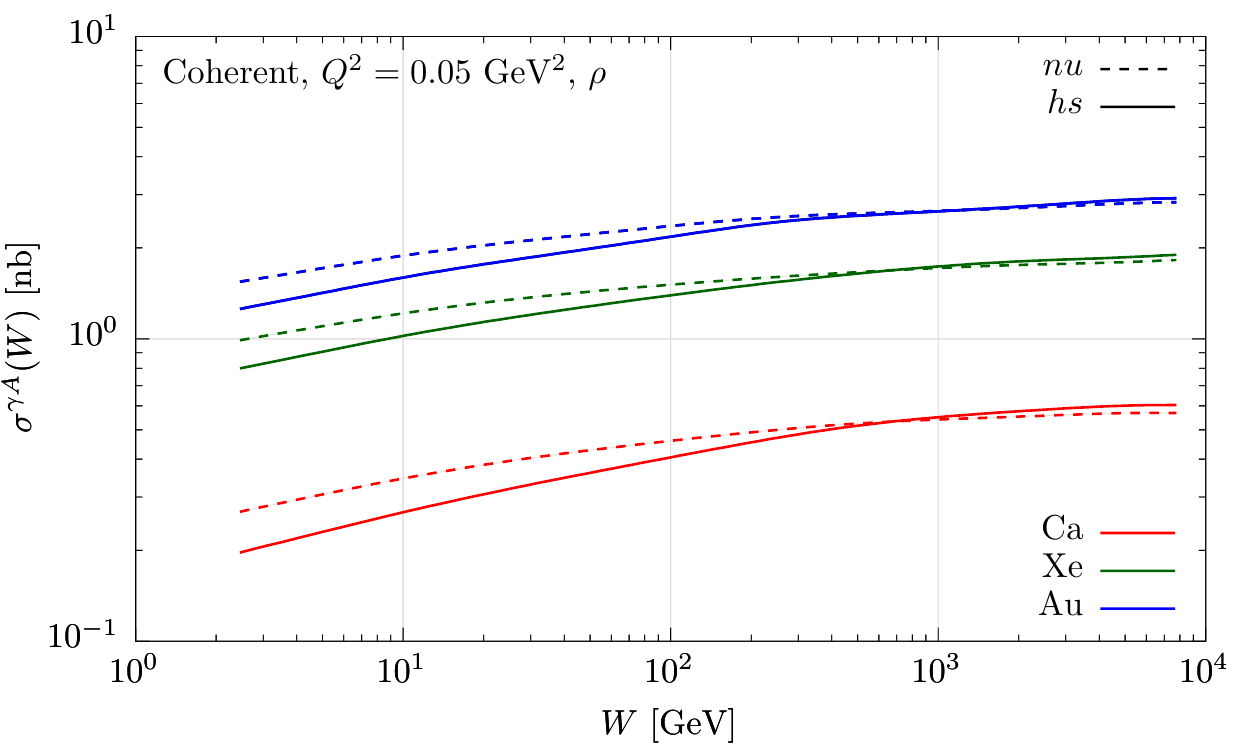}}%
\qquad
\subfigure{%
\includegraphics[width=0.4\textwidth]{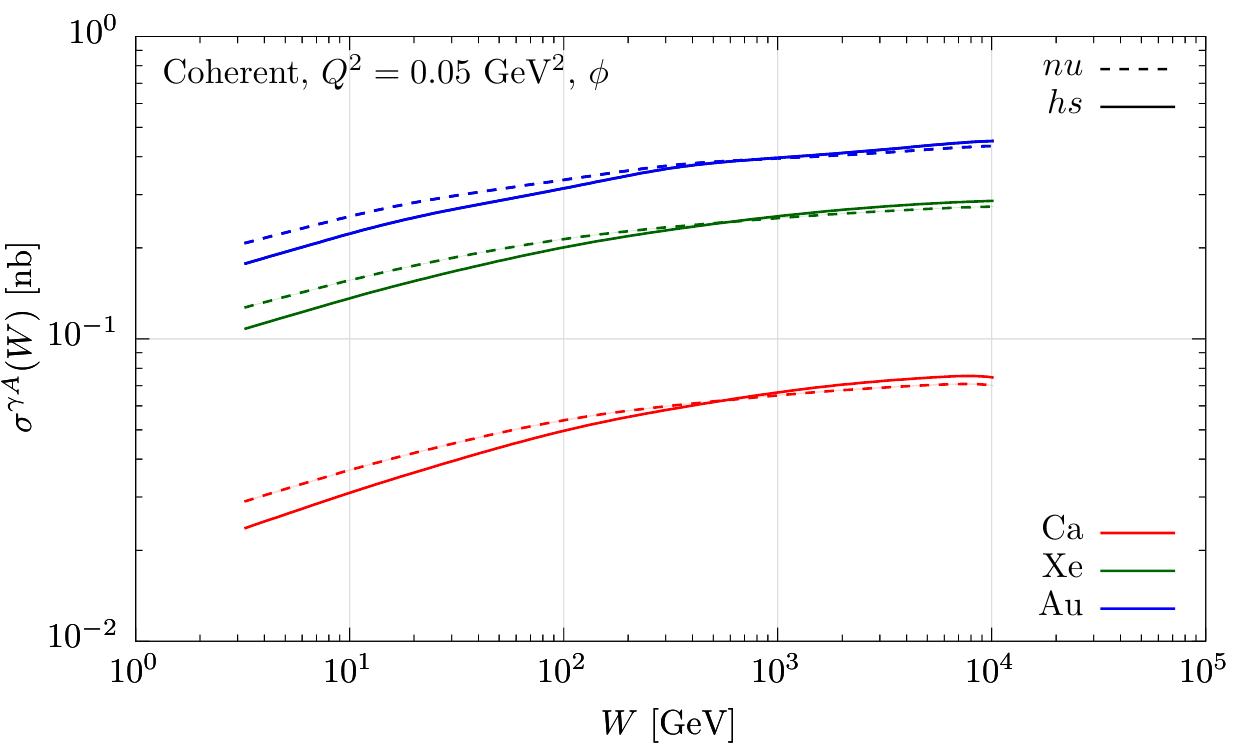}}\\%
\subfigure{%
\includegraphics[width=0.4\textwidth]{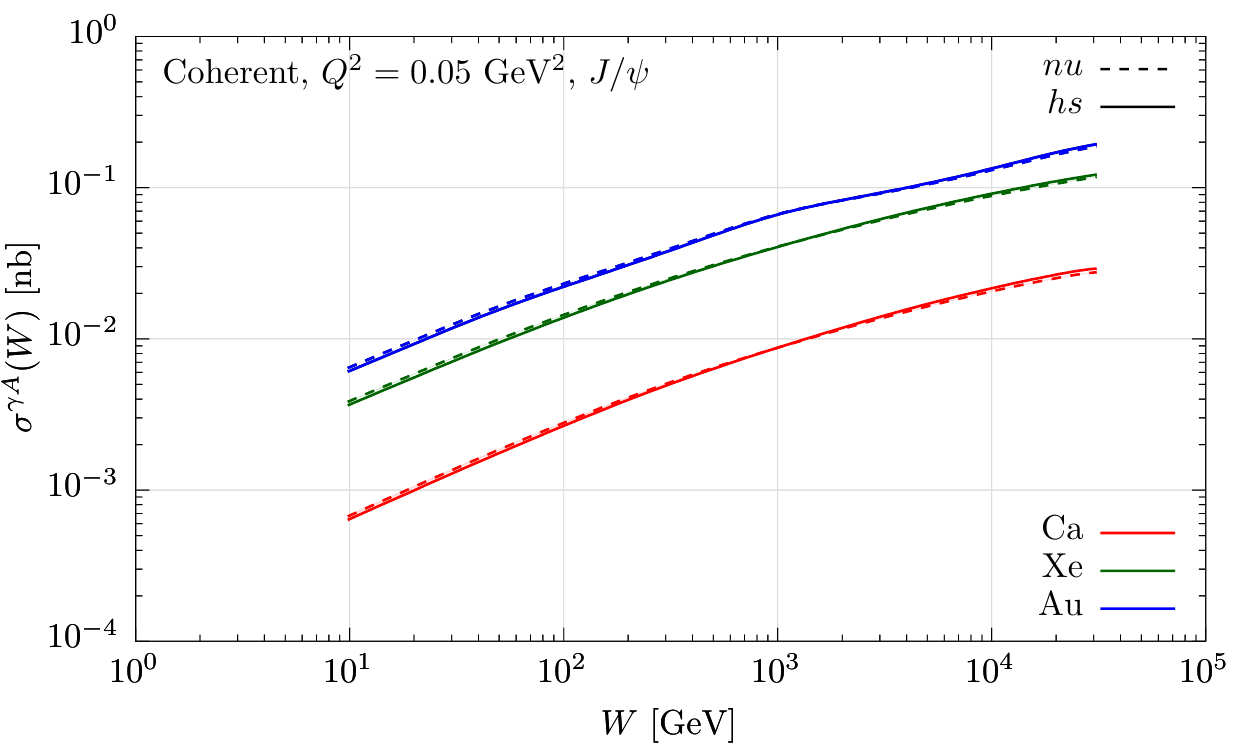}}%
\qquad
\subfigure{%
\includegraphics[width=0.4\textwidth]{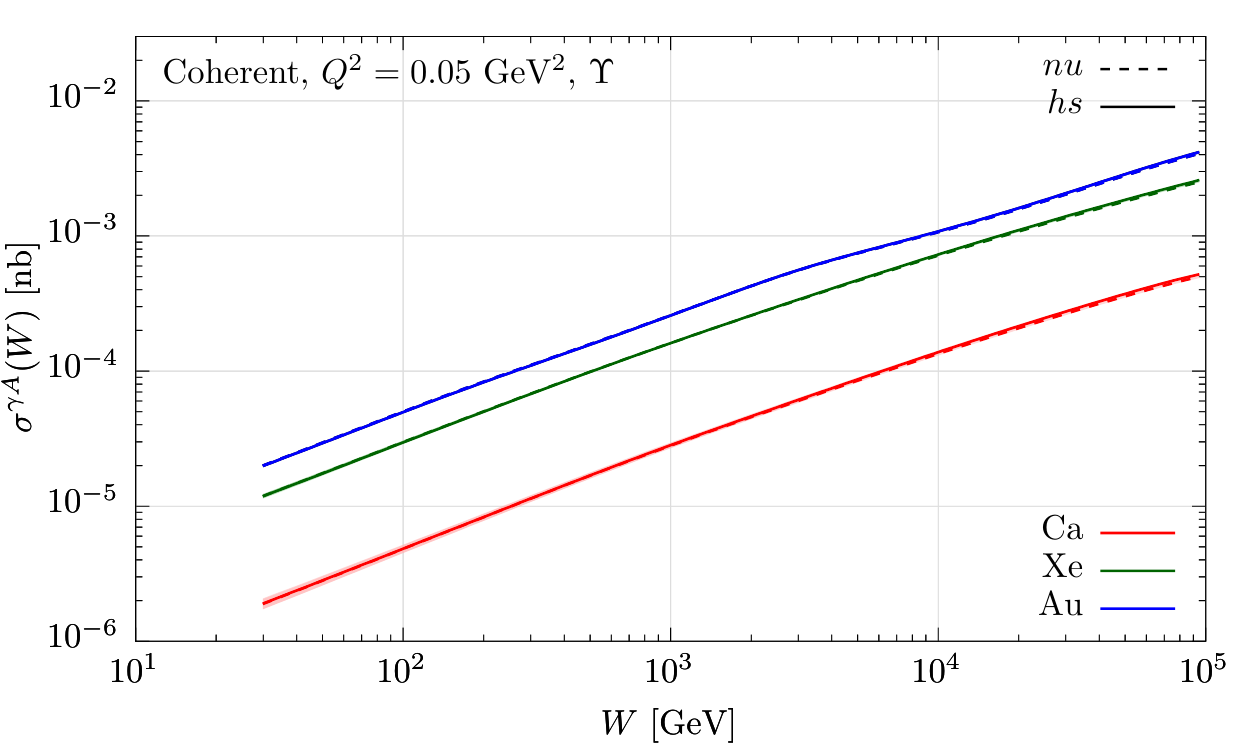}}%
\caption{Predictions for the energy dependence of the cross sections for the coherent exclusive vector meson production considering different values of the atomic number and $Q^2 = 0.05$ GeV$^2$. The solid (dashed) lines correspond to the predictions of the $hs$ ($nu$) model for the nuclear profile.}
\label{fig:xseccoh-nuclei}%
\end{figure}

In Fig. \ref{fig:xseccoh-energy} we present our predictions for the energy dependence of the coherent cross sections for different values of the photon virtuality $Q^2$ considering an $eAu$ collision. The dashed lines correspond to the predictions obtained assuming that nuclear profile is  made up of nucleons, denoted $nu$, and the solid lines to the nuclear profile made up of hot spots, $hs$. In agreement with previous studies \cite{vmprc,Caldwell,Lappi_inc,Toll,Lappi_bal,diego,Mantysaari:2017slo}, we have that the cross sections increase with energy and decrease with $Q^2$. Moreover, we have that for a fixed $Q^2$ the increasing with $W$ is steeper for heavier vector mesons, which is directly associated to the fact that for these mesons the cross section is dominated by smaller color dipoles and, therefore, the impact of non - linear effects in the QCD dynamics is reduced, independently of the value of the photon virtuality.
In contrast, in the case of the light vector mesons, we have that at small photon virtualities the main contribution comes from  large size dipoles, with the dynamics being determined by the gluon saturation effects. When the photon virtuality increases, the contribution of smaller dipoles becomes larger, reducing the contribution of non - linear effects. As a consequence, the cross sections for the light vector meson production become steeper with the energy at larger values of $Q^2$. Such behaviour is observed in the upper panels of Fig. \ref{fig:xseccoh-energy}. Regarding the impact of the modeling of $T_A(\rb)$, we have that the $nu$ and $hs$ predictions are almost identical, which is expected by the fact that the coherent cross section probes the average over configurations of the nuclear wave function. Such behaviour is also verified when we consider the electron - ion collisions for different nuclei, as presented in Fig. \ref{fig:xseccoh-nuclei}.

\begin{figure}%
\centering
\subfigure{%
\includegraphics[width=0.4\textwidth]{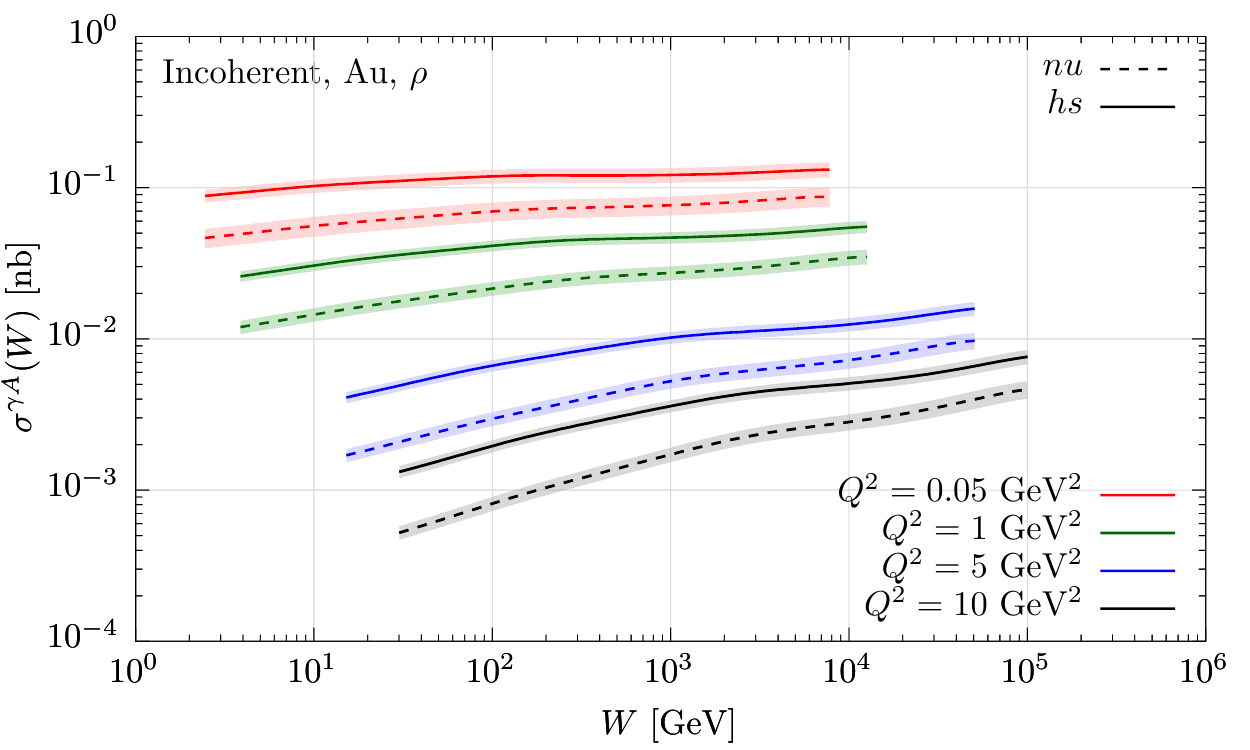}}%
\qquad
\subfigure{%
\includegraphics[width=0.4\textwidth]{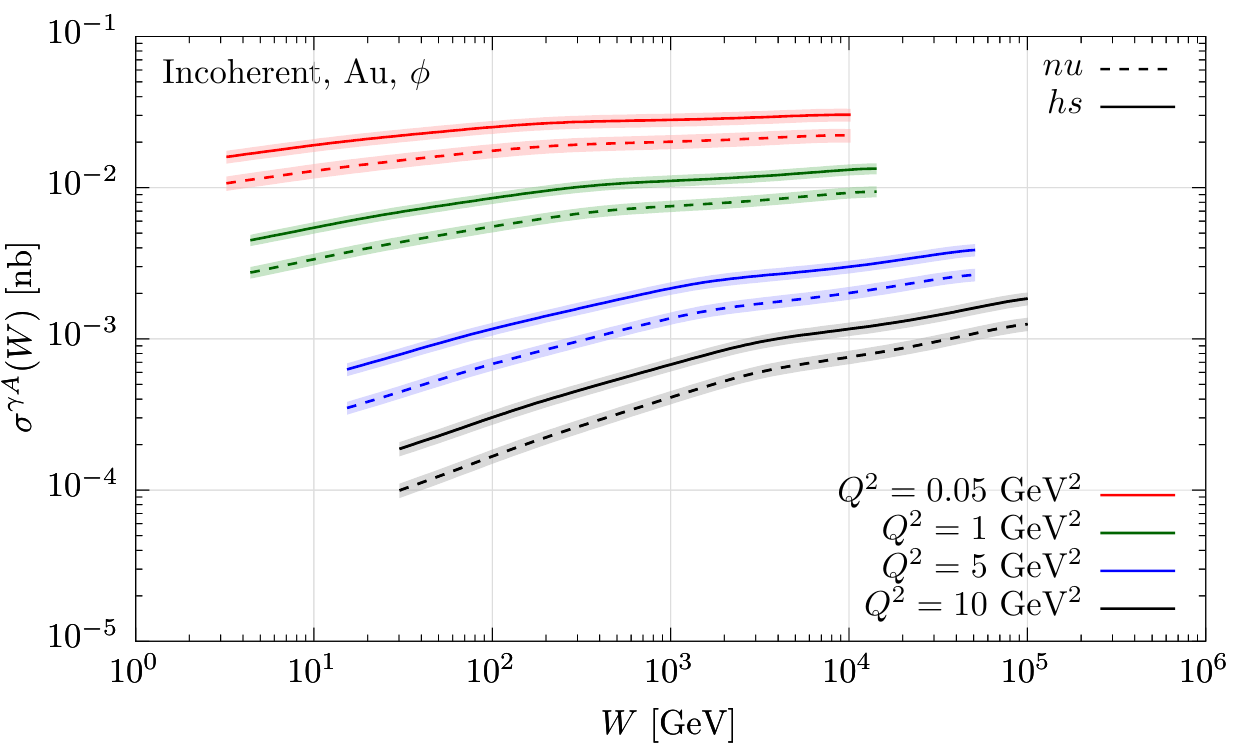}}\\%
\subfigure{%
\includegraphics[width=0.4\textwidth]{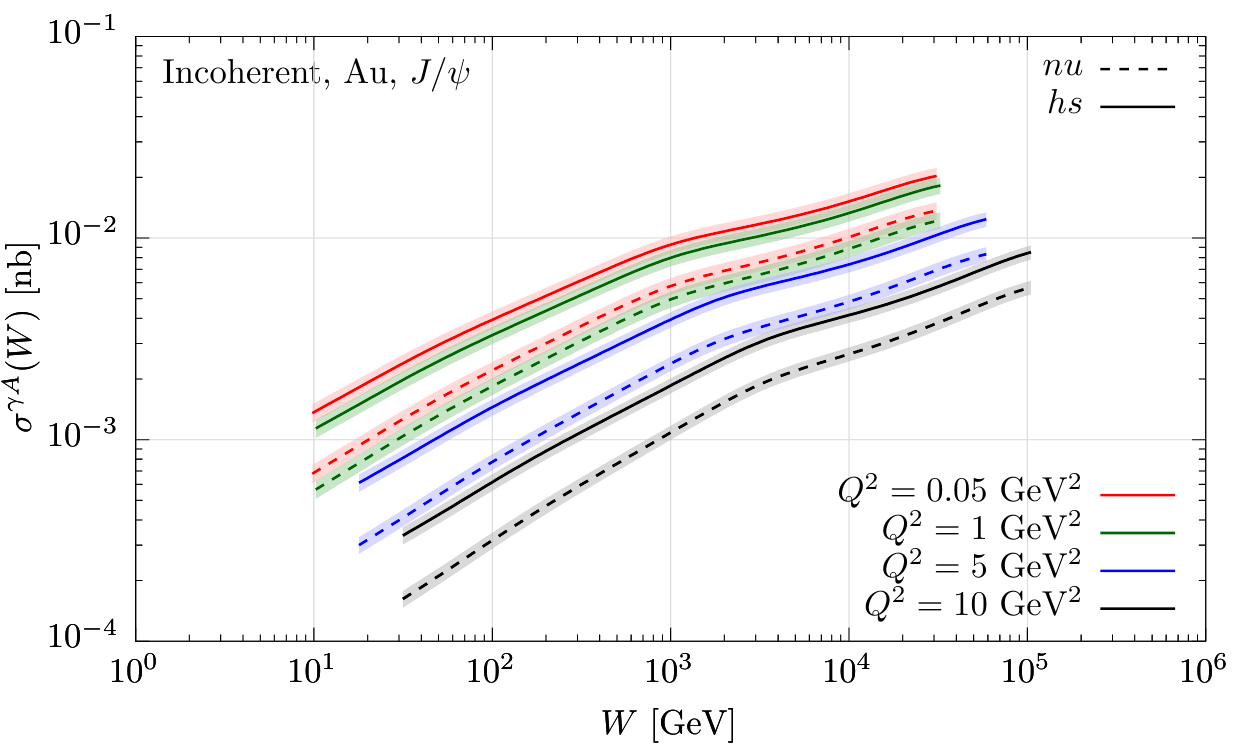}}
\qquad
\subfigure{%
\includegraphics[width=0.4\textwidth]{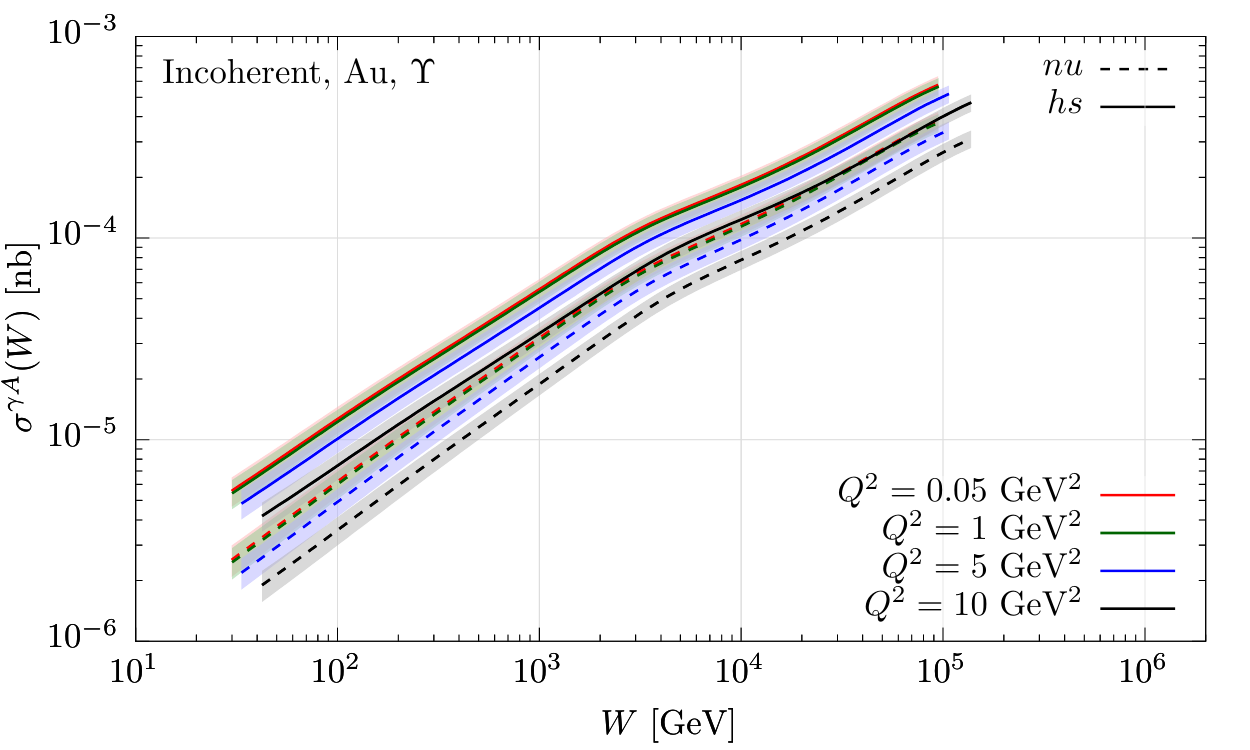}}
\caption{Predictions for incoherent exclusive vector meson production considering a Gold target as a function of energy $W$ for different values of $Q^2$. The solid (dashed) lines correspond to the predictions of the $hs$ ($nu$) model for the nuclear profile.}
\label{fig:xsecinc-energy}%
\end{figure}

\begin{figure}%
\centering
\subfigure{%
\includegraphics[width=0.4\textwidth]{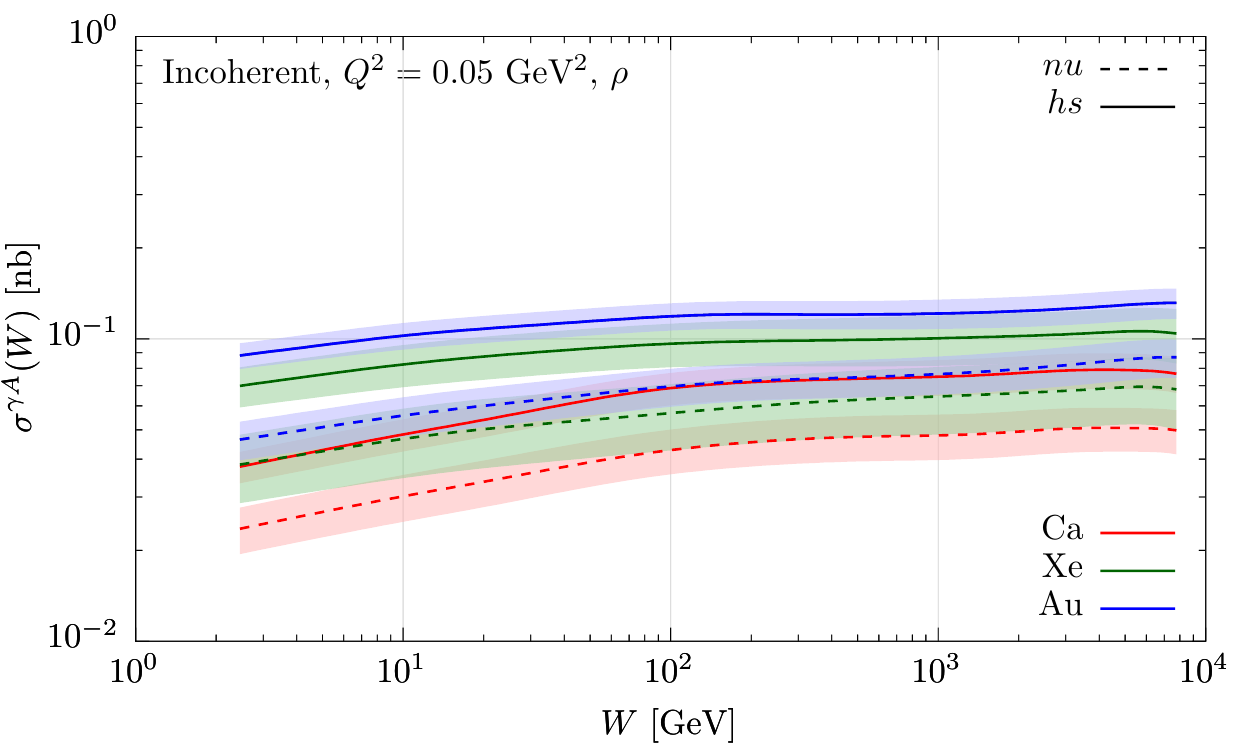}}%
\qquad
\subfigure{%
\includegraphics[width=0.4\textwidth]{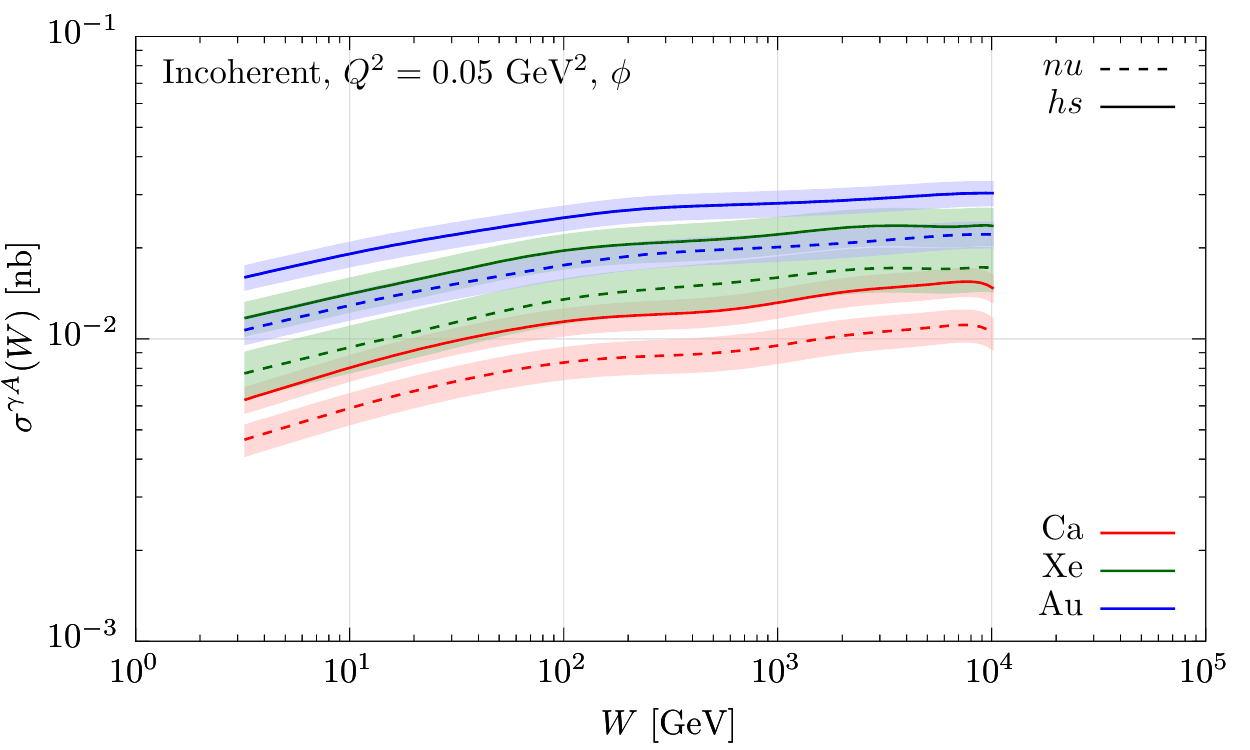}}\\%
\subfigure{%
\includegraphics[width=0.4\textwidth]{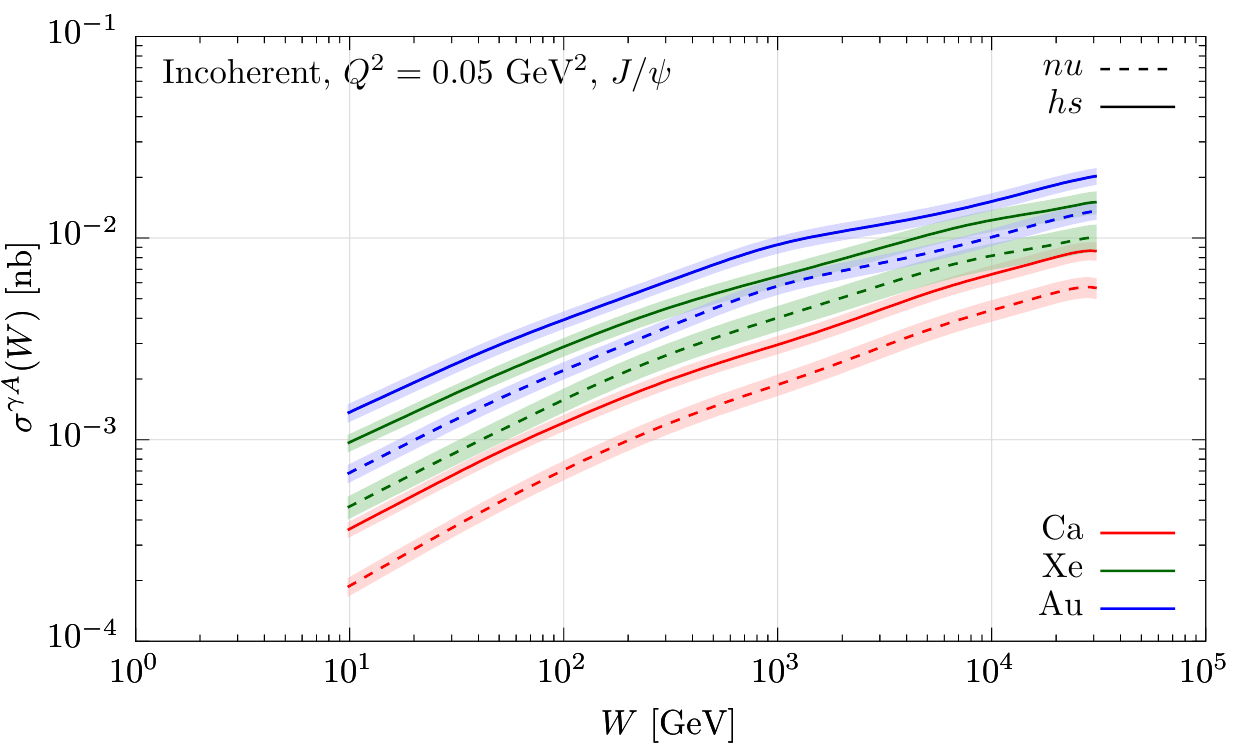}}
\qquad
\subfigure{%
\includegraphics[width=0.4\textwidth]{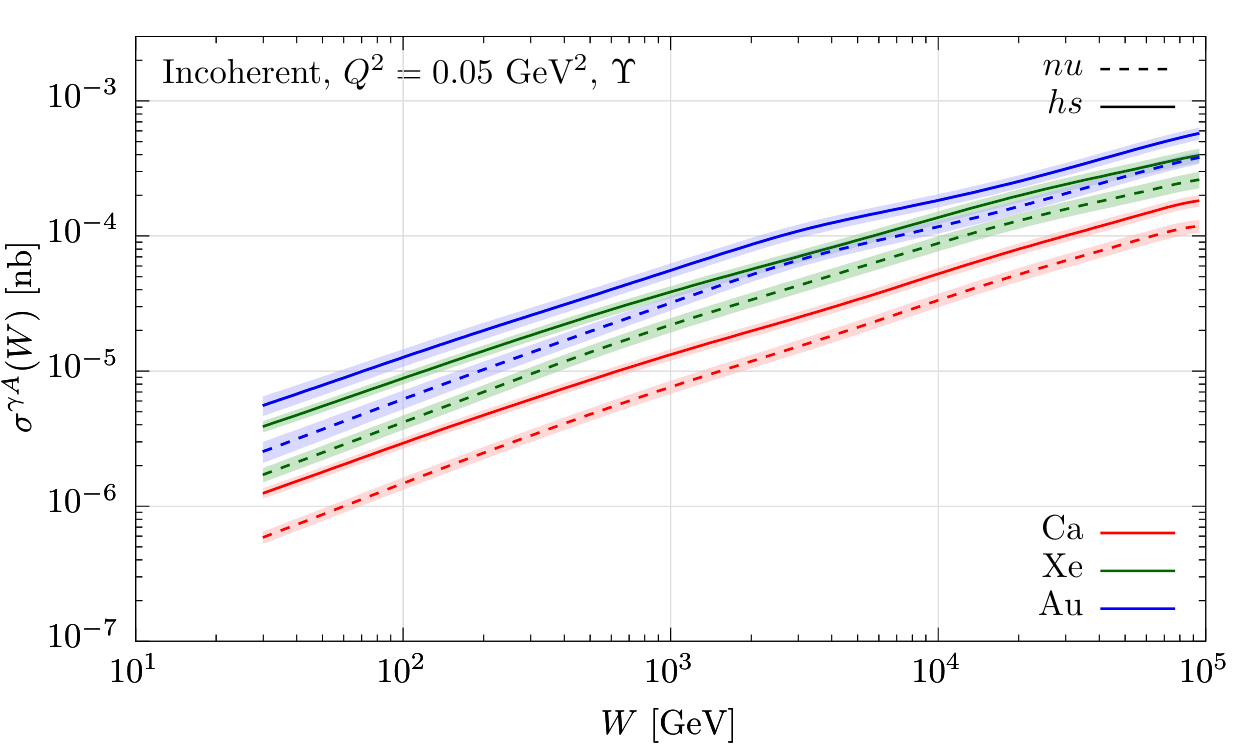}}%
\caption{Predictions for the energy dependence of the cross sections for the incoherent exclusive vector meson production considering different values of the atomic number and $Q^2 = 0.05$ GeV$^2$. The solid (dashed) lines correspond to the predictions of the $hs$ ($nu$) model for the nuclear profile.}
\label{fig:xsecinc-nuclei}%
\end{figure}

\begin{figure}%
\centering
\subfigure{%
\includegraphics[width=0.4\textwidth]{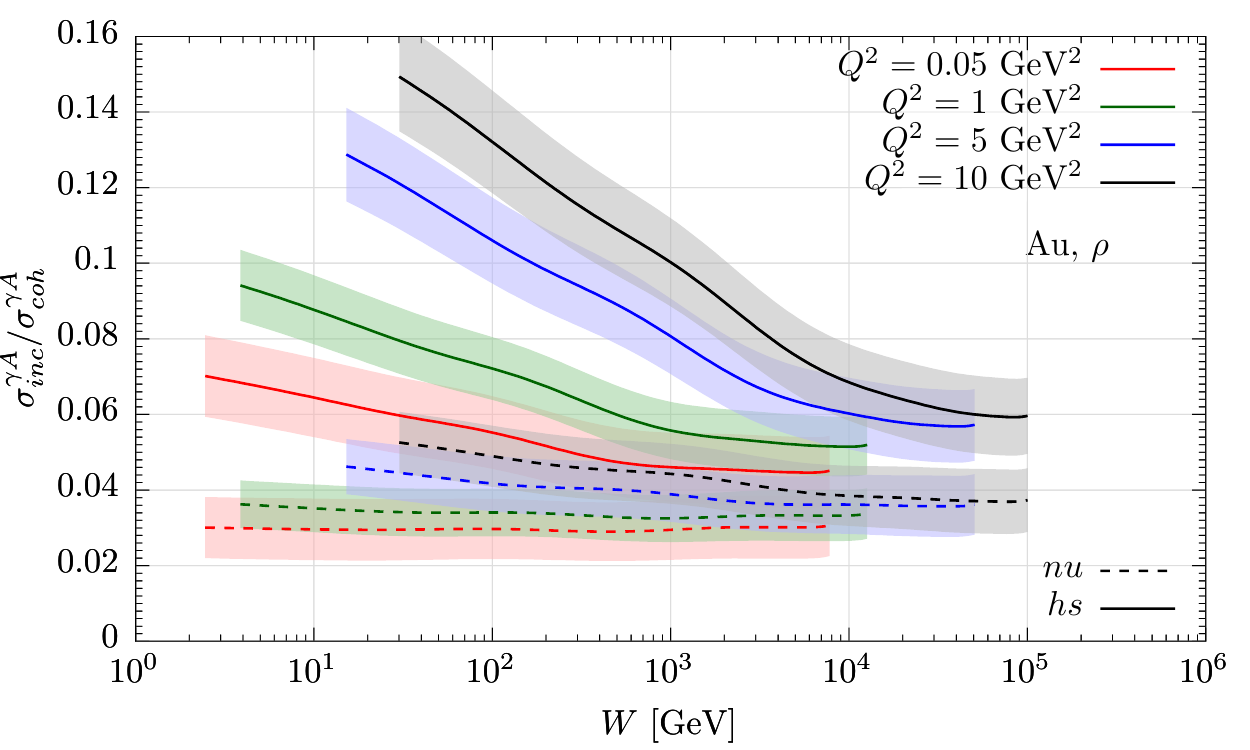}}%
\qquad
\subfigure{%
\includegraphics[width=0.4\textwidth]{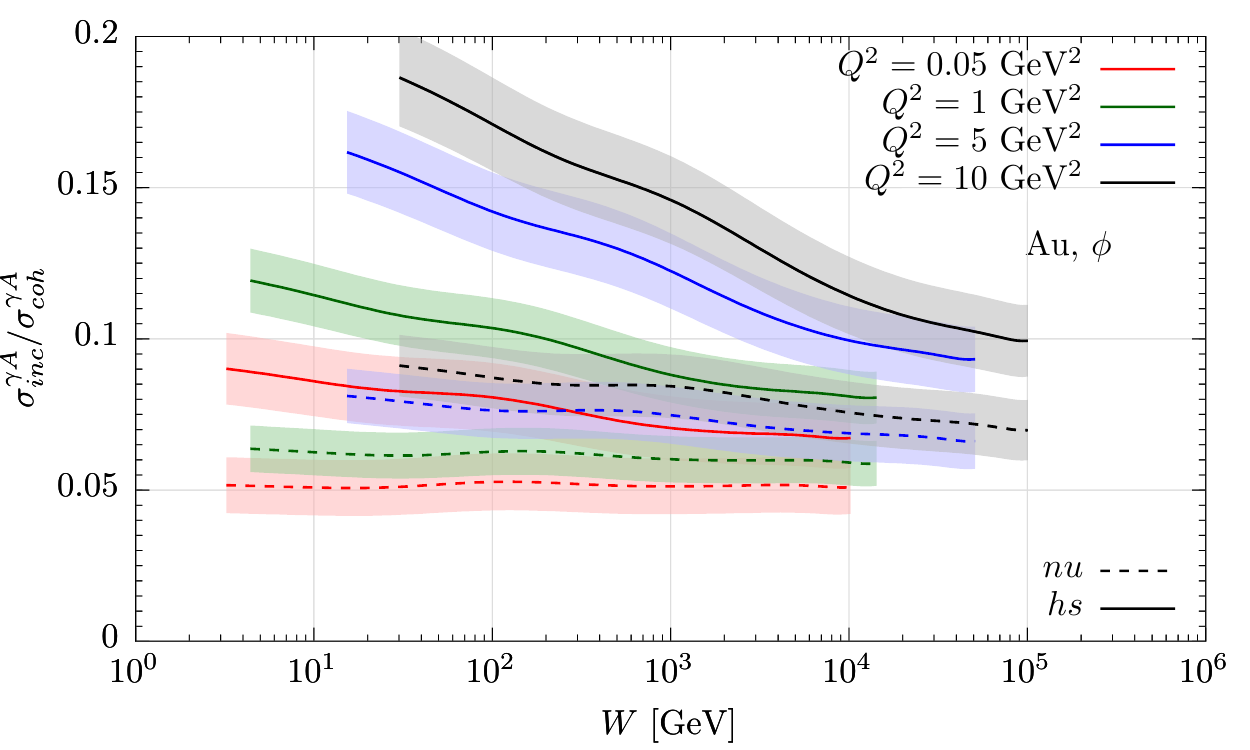}}\\
\subfigure{%
\includegraphics[width=0.4\textwidth]{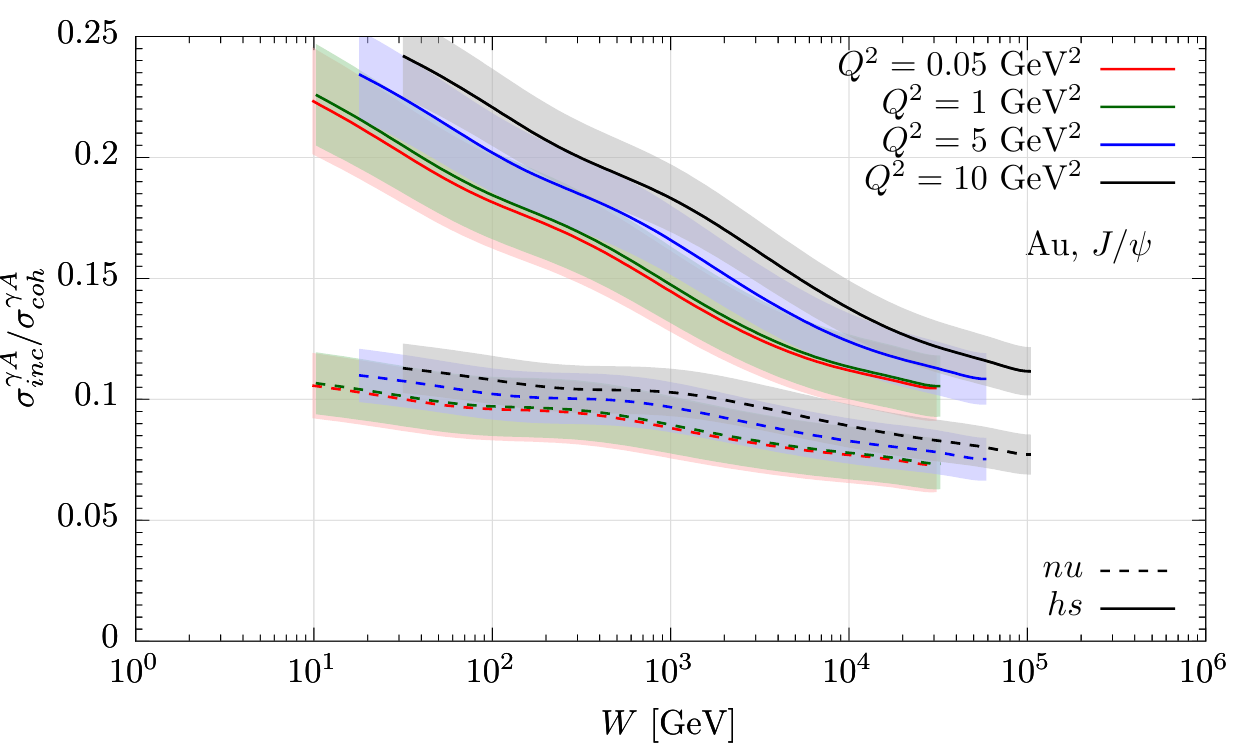}}%
\qquad
\subfigure{%
\includegraphics[width=0.4\textwidth]{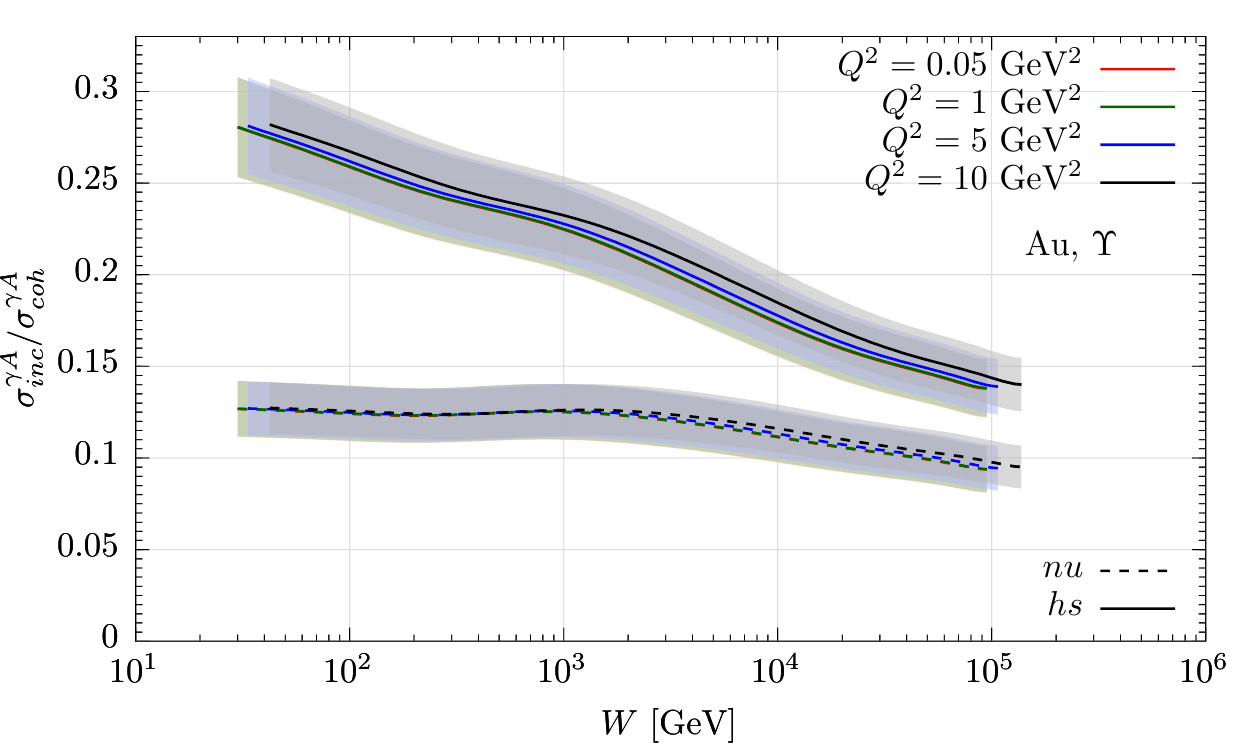}}\\
\caption{Predictions for the ratio between the incoherent and coherent cross sections as  as a function of energy $W$ for different values of $Q^2$ considering $eAu$ collisions. The solid (dashed) lines correspond to the predictions of the $hs$ ($nu$) model for the nuclear profile.}
\label{fig:ratio-energy}%
\end{figure}

\begin{figure}%
\centering
\subfigure{%
\includegraphics[width=0.4\textwidth]{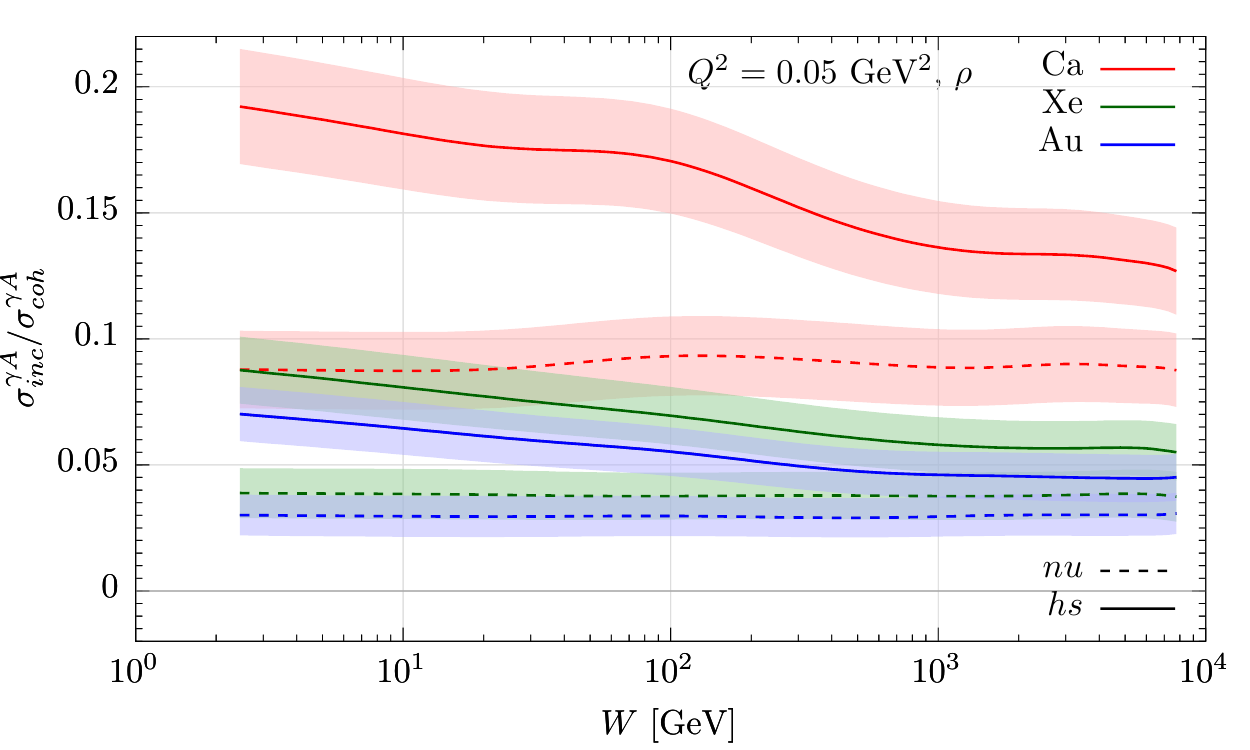}}%
\qquad
\subfigure{%
\includegraphics[width=0.4\textwidth]{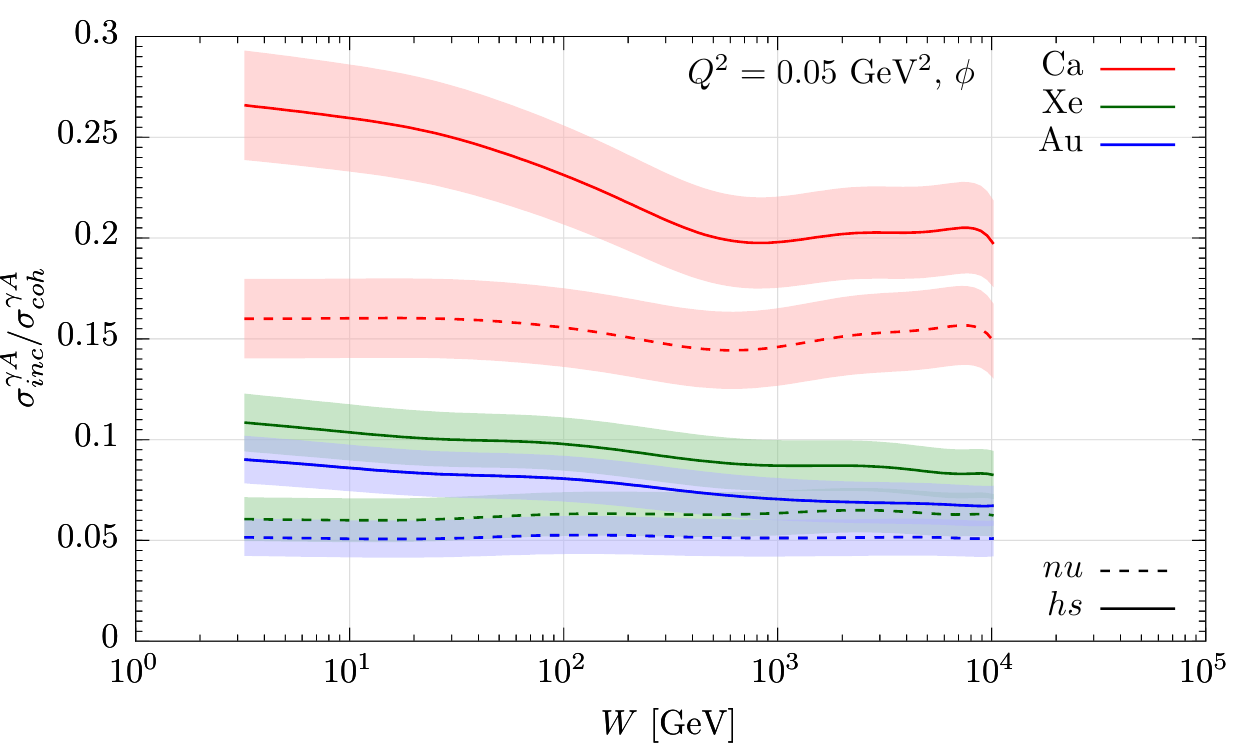}} \\
\subfigure{%
\includegraphics[width=0.4\textwidth]{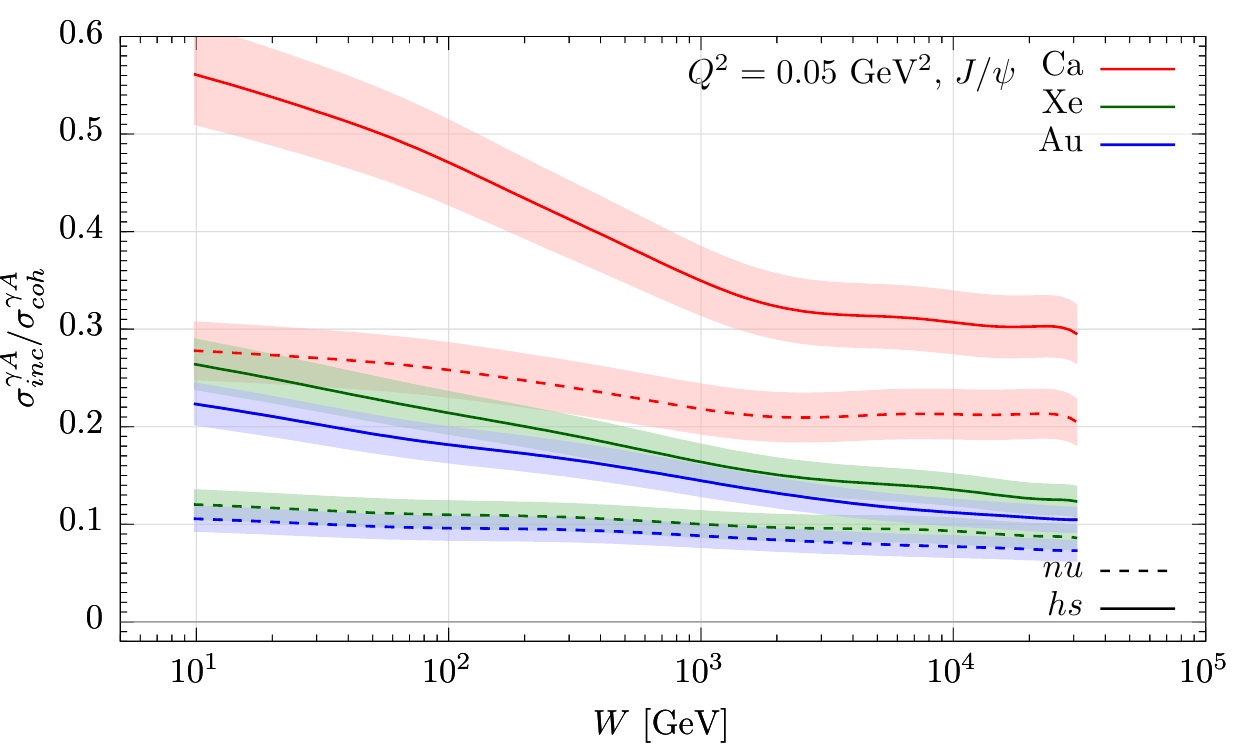}}%
\qquad
\subfigure{%
\includegraphics[width=0.4\textwidth]{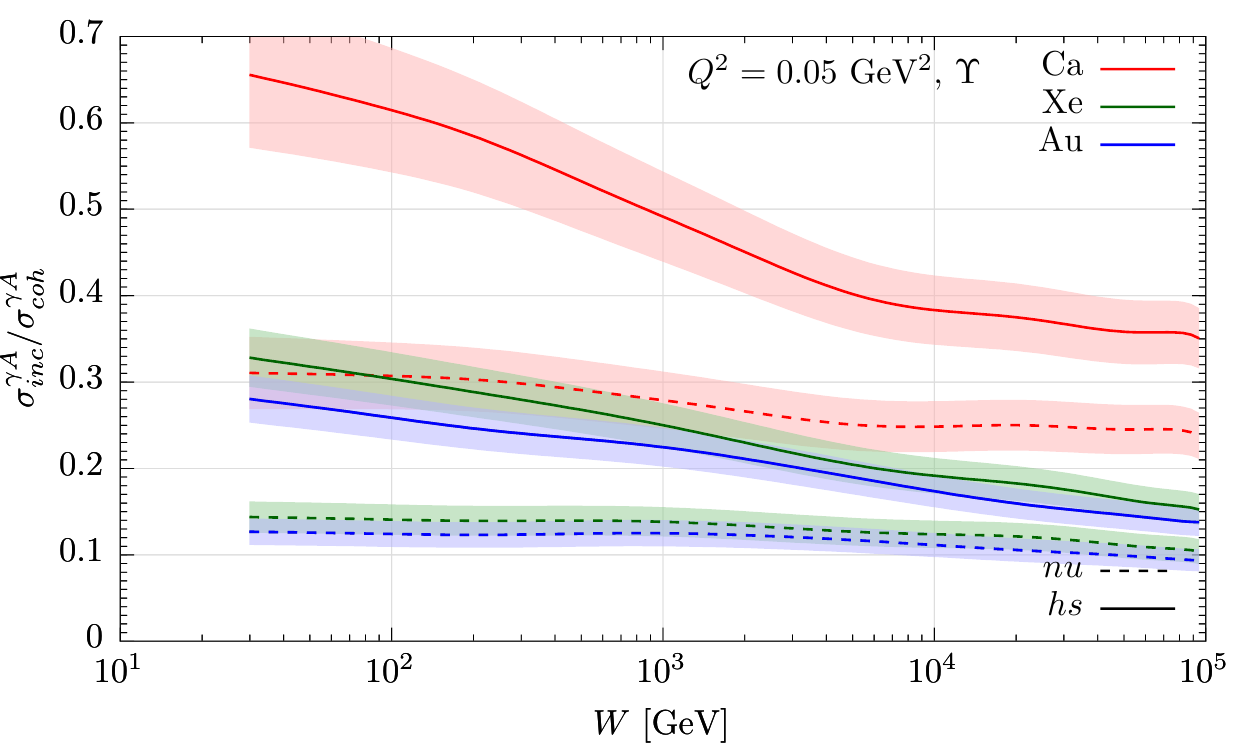}}%
\caption{Predictions for the ratio between the incoherent and coherent cross sections as  as a function of energy $W$ for different atomic nuclei and $Q^2 = 0.05$ GeV$^2$. The solid (dashed) lines correspond to the predictions of the $hs$ ($nu$) model for the nuclear profile.}
\label{fig:ratio-nuclei}%
\end{figure}

In Figs. \ref{fig:xsecinc-energy} and \ref{fig:xsecinc-nuclei} we present our predictions for the exclusive vector meson production in incoherent interactions. Similarly to the coherent case, the cross sections  decrease with $Q^2$ and increase with the energy, with the increase being dependent on the vector meson considered. However, for the incoherent production, the predictions are sensitive to the description of the nuclear profile. We have that the $hs$ model implies larger values for the incoherent cross sections, with the enhancement in comparison to the $nu$ model being present for all values of $Q^2$ and atomic number. In order to quantify this impact and reduce the systematic uncertainty in our calculations, we will estimate the ratio between the incoherent and coherent cross sections. Our predictions for the energy, $Q^2$ and atomic number dependencies of this ratio are presented in Figs.
\ref{fig:ratio-energy} and \ref{fig:ratio-nuclei}. We have that both models predict that the ratio decreases at smaller values of the photon virtuality and larger nuclei. The difference between the $nu$ and $hs$ predictions is larger when the photon virtuality is increased, especially for lighter vector mesons. Such a result is associated with the fact that $x \propto Q^2 + M^2$ for a fixed energy. Consequently, at larger $Q^2$ we are probing larger values of $x$, where the number of hot - spots is smaller, which implies a larger variance in the event - by - event configurations. For heavy mesons, the predictions become sensitive to $Q^2$ only when $Q^2 \gg M^2$.
The dependence on $A$ of the ratio is connected to the fact that for lighter nuclei we have a smaller number of nucleons in longitudinal coordinate $z$, which induces more inter-nucleon (inter-nucleon-hot-spots) space and, consequently, increases the variance between configurations. Finally, in contrast with the $nu$ model, the $hs$ model predicts that the ratio is strongly dependent on the energy, with the difference between the predictions being smaller at larger energies. This behaviour is directly associated to the energy dependence of the number of hot - spots. With the increasing of the energy $W$, and, consequently, decreasing of $x$, one has a higher number of hot spots as given by Eq. \eqref{eq:Nhs}. The growing number of hot-spot tends to fill up the nucleon, and both models approach each other. Our results indicate that future experimental analysis of this ratio can be useful to constrain the description of the nuclear profile and, in particular, to probe the presence of hot-spots inside the nucleons.

\begin{figure}%
\centering
\subfigure{%
\includegraphics[width=0.4\textwidth]{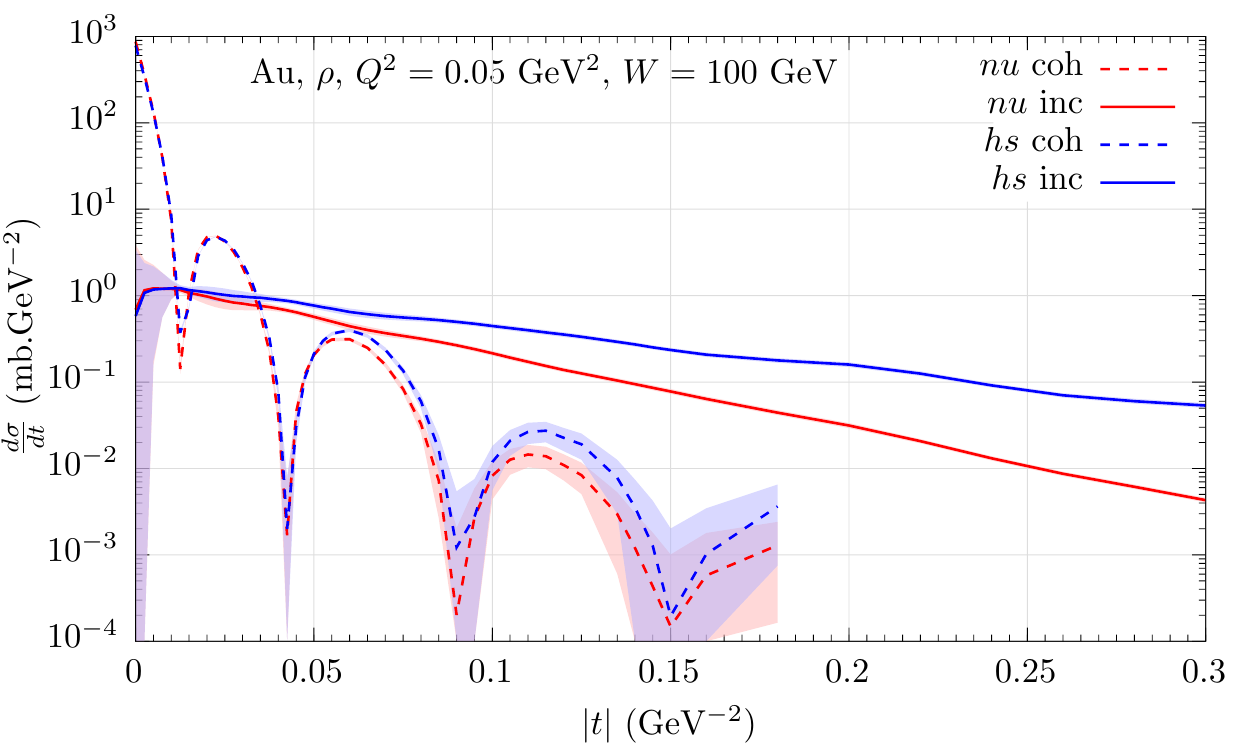}}%
\qquad
\subfigure{%
\includegraphics[width=0.4\textwidth]{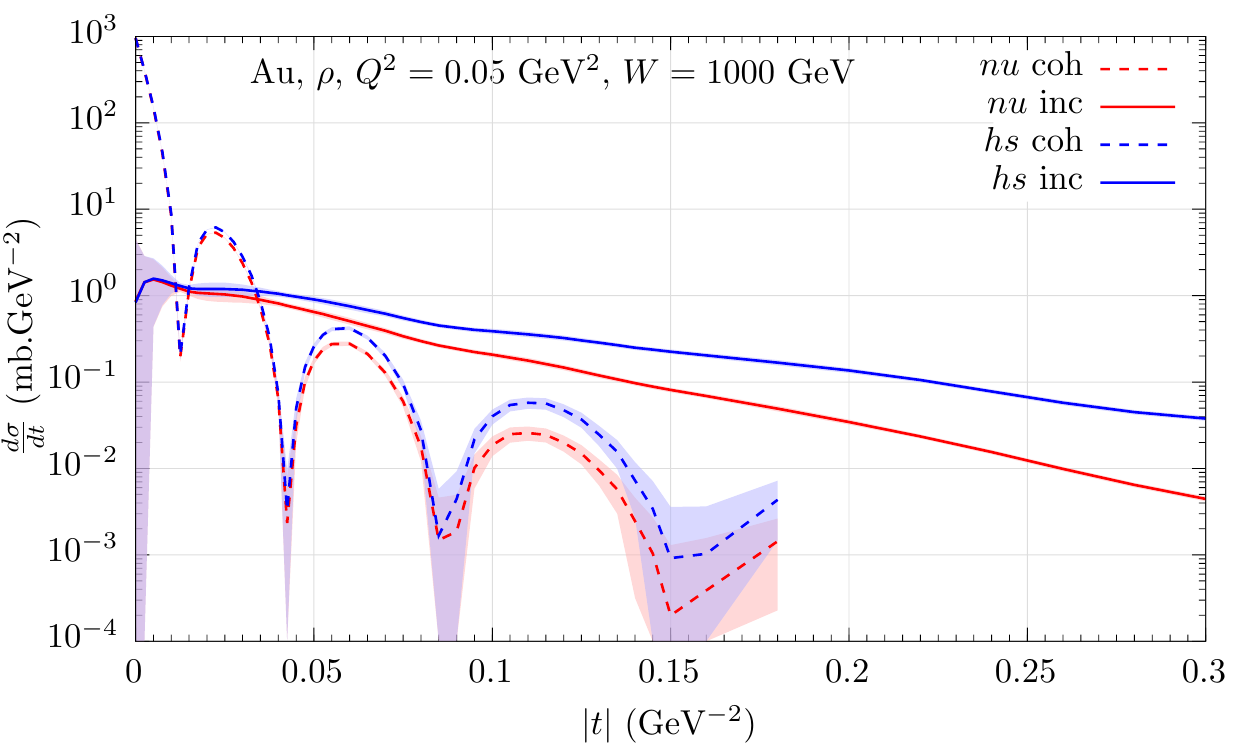}}\\
\subfigure{%
\includegraphics[width=0.4\textwidth]{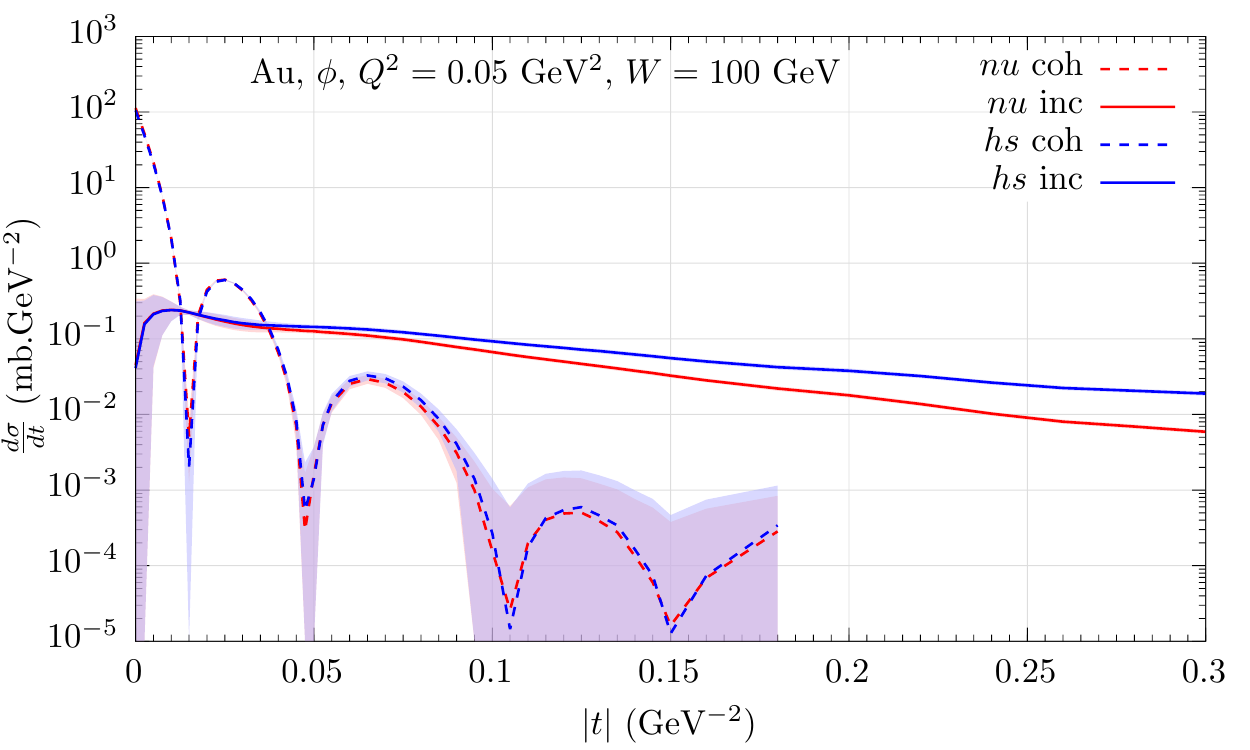}}%
\qquad
\subfigure{%
\includegraphics[width=0.4\textwidth]{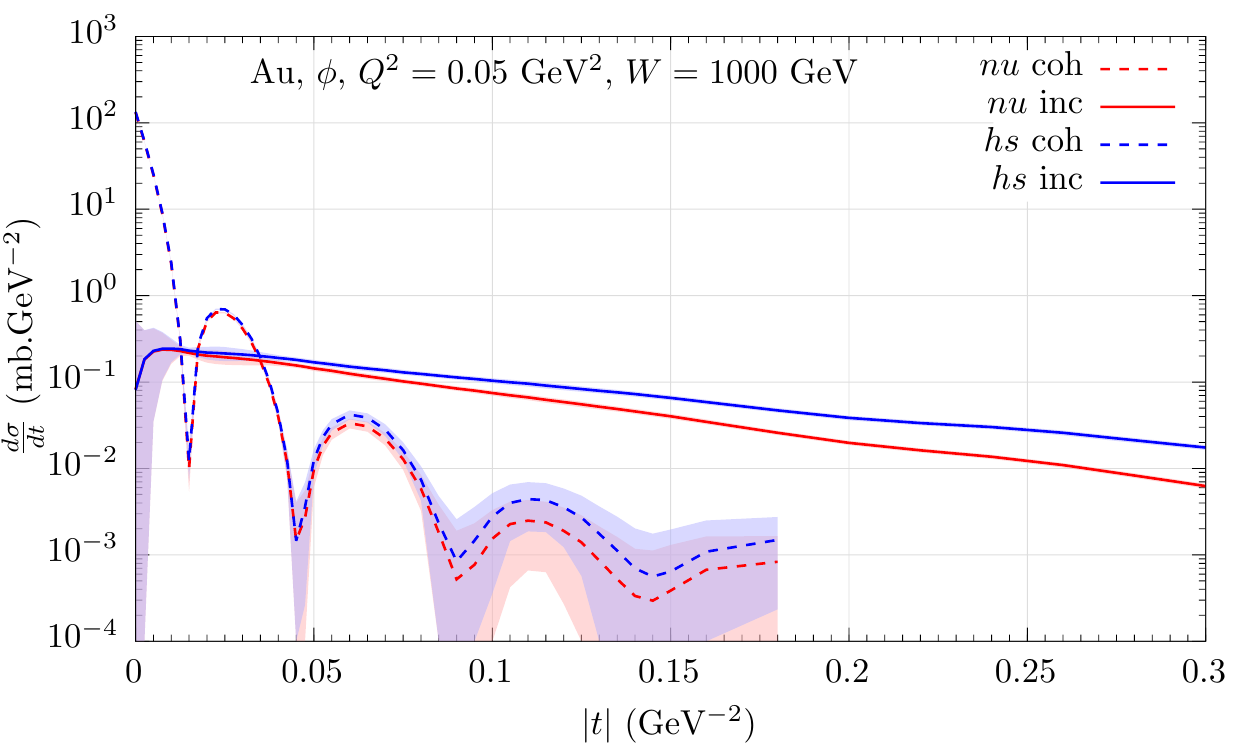}}\\
\subfigure{%
\includegraphics[width=0.4\textwidth]{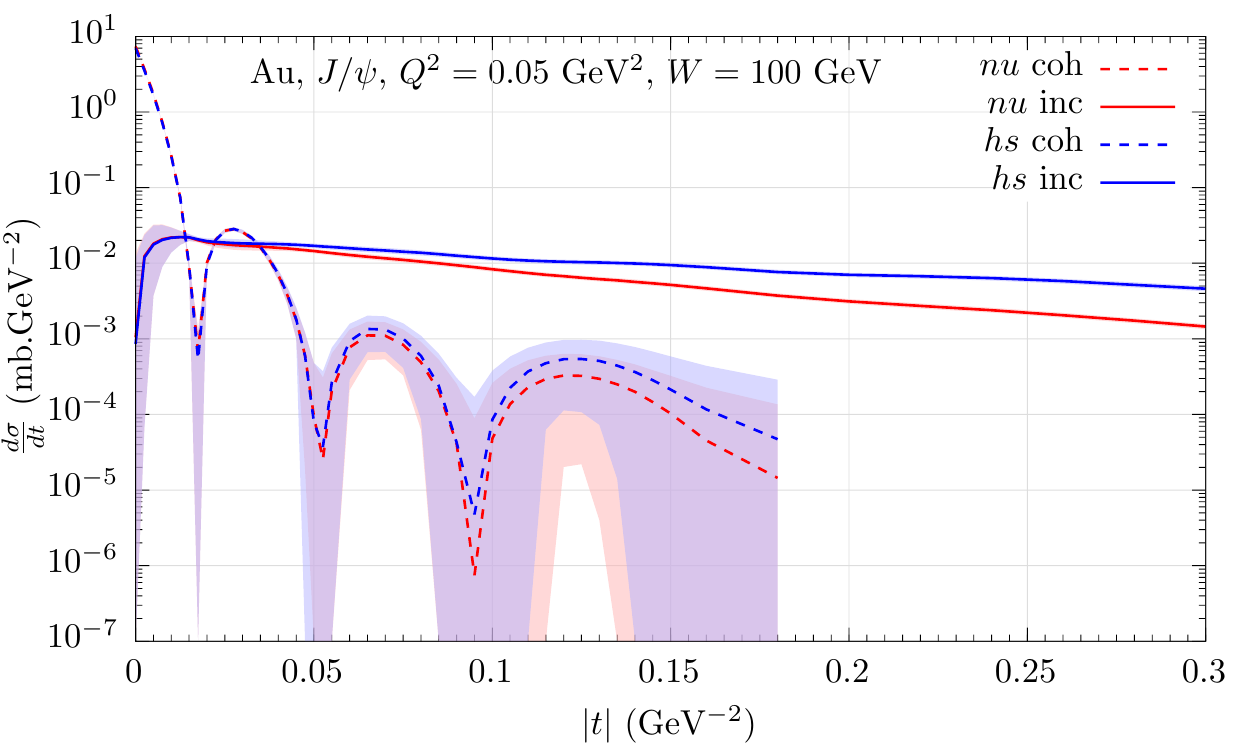}}%
\qquad
\subfigure{%
\includegraphics[width=0.4\textwidth]{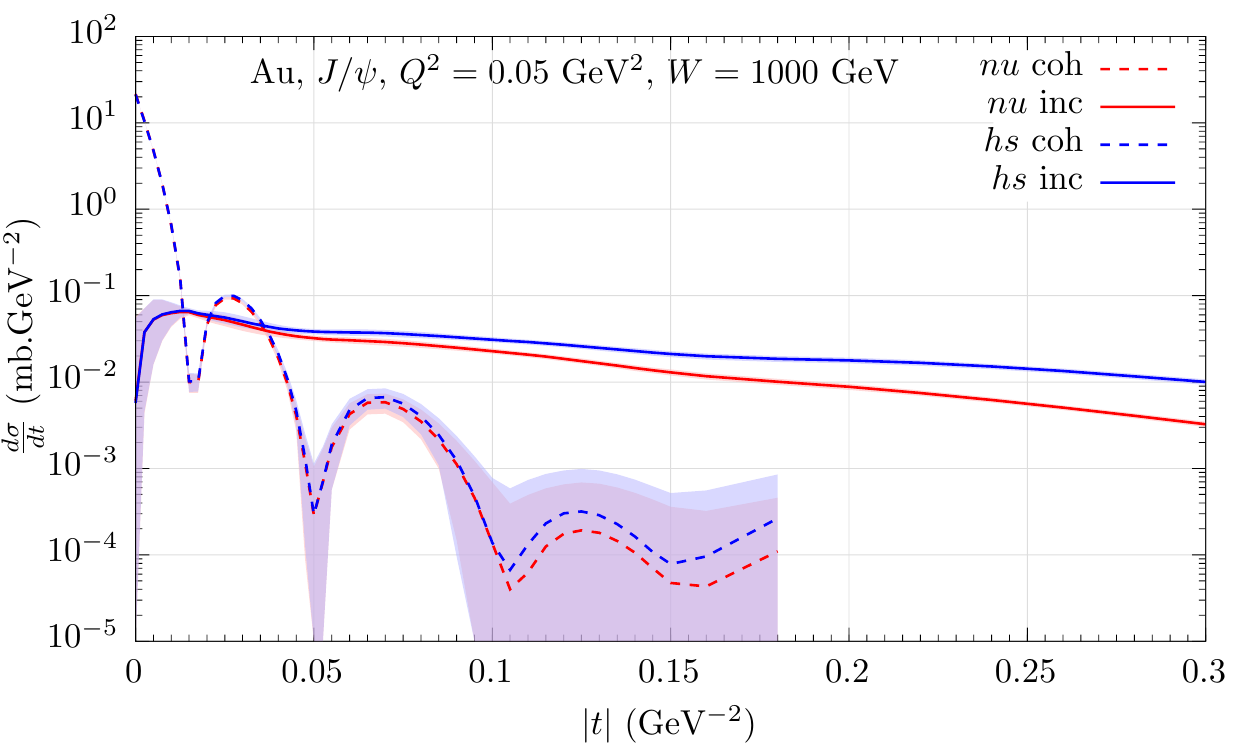}}\\
\subfigure{%
\includegraphics[width=0.4\textwidth]{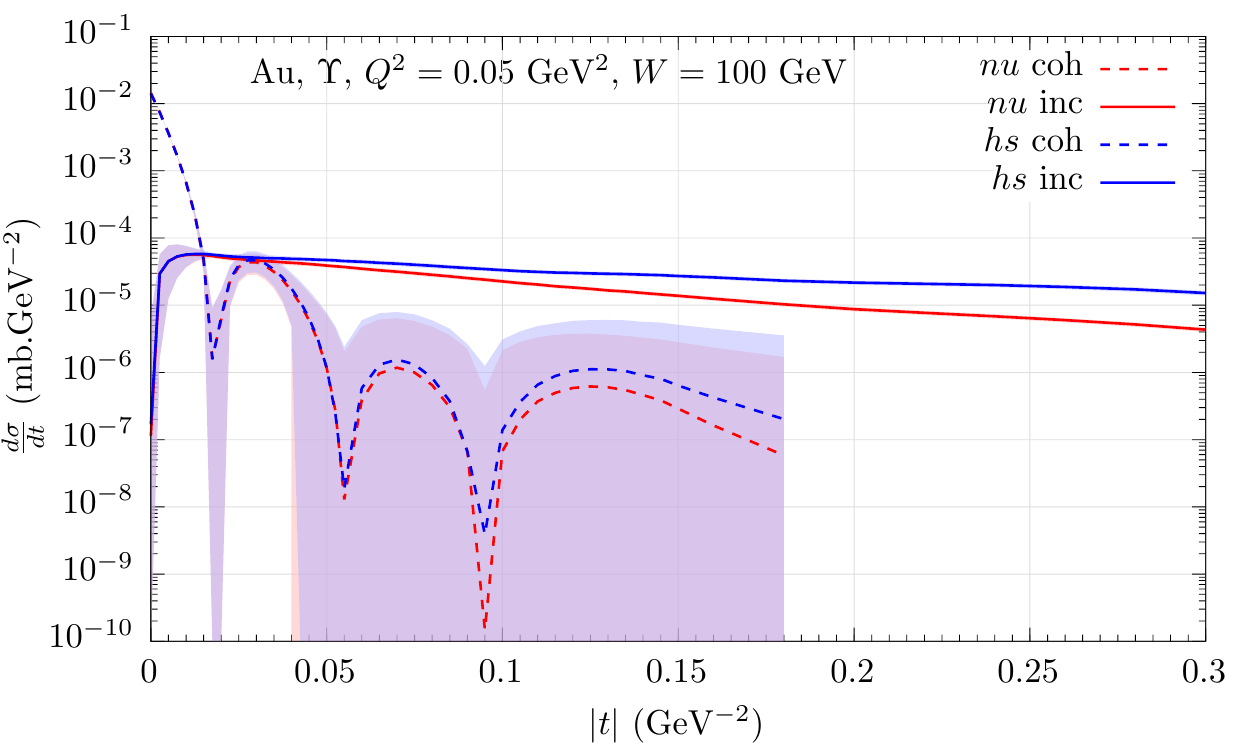}}%
\qquad
\subfigure{%
\includegraphics[width=0.4\textwidth]{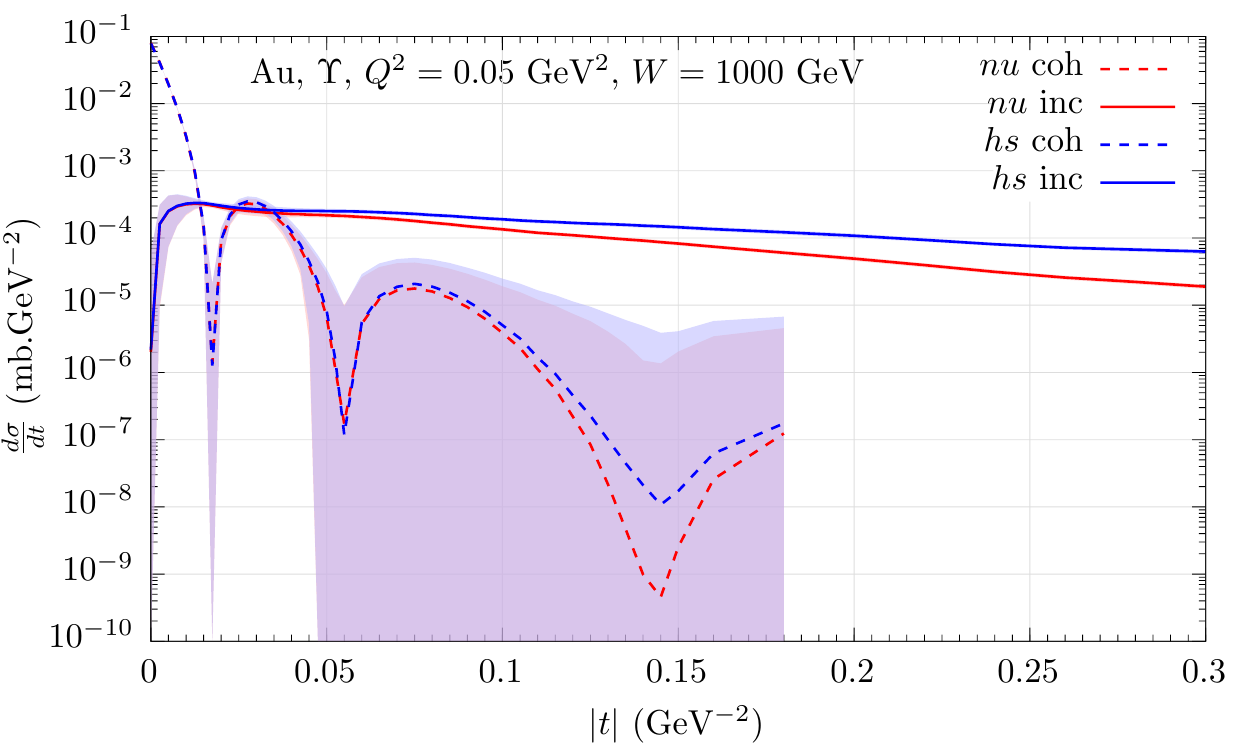}}
\caption{Predictions for the $t$ - dependence of the coherent and incoherent cross sections for $\gamma^*Au$ interactions at  $W=100$~GeV (left) and $W=1000$~GeV (right). The solid (dashed) lines correspond to the predictions of the $hs$ ($nu$) model for the nuclear profile.}
\label{fig:tdist2}%
\end{figure}

Finally, in Fig. \ref{fig:tdist2} we present our predictions for the $t$ - distributions considering the exclusive vector meson production in coherent and incoherent interactions. We assume $Q^2 = 0.05$ GeV$^2$ and consider  $W = 100$ and 1000 GeV. The results are presented in the left and right panels, respectively. The incoherent (coherent) predictions are represented by solid (dashed) lines.
The coherent cross sections clearly exhibit the typical diffractive pattern  and are  characterized by a sharp forward diffraction peak. In contrast, the incoherent cross sections are characterized by a $t$ - dependence similar to that observed in the vector meson $\rho$ production off free nucleons. One has that the incoherent processes dominate at large - $|t|$  and the coherent ones at small values of the momentum transfer. Such behaviour is expected: with the increasing of the momentum kick given to the nucleus the probability that  it  breaks up becomes larger. As a consequence, the vector meson production at large - $|t|$ is dominated by incoherent processes.
The results  presented in Fig. \ref{fig:tdist2} are in accordance with those presented previously in Refs. \cite{Cepila:2017nef,Mantysaari:2017dwh} for the $J/\Psi$ case, but are here extended to different vector mesons and energies. One has that the difference between incoherent cross sections for $hs$ and $nu$ models increases with increasing momentum transfer $t$.
The increasing difference can also be observed for the coherent cross section with increasing $t$. However, this plays a very small role in the integrated cross section because of small absolute contributions. Focusing on the comparison between energy $W=100$~GeV (left figures) and 1000~GeV (right figures), we do not observe significant difference except the higher difference between $nu$ and $hs$ model for the coherent cross section at large $t$.

%
%

\section{Conclusions}
\label{sec:sum}

The study of exclusive processes in deep inelastic scattering (DIS) electron - nucleus processes probe the QCD dynamics at high
energies,  driven by the gluon content of the nucleus, which is strongly subject to non-linear effects (parton saturation). Our goal in this paper was to present a comprehensive analysis of the energy, virtuality, nuclear mass number and transverse momentum dependencies of the cross sections for the vector meson production in the kinematical range which could be accessed in future electron - ion colliders.
In particular, our focus was in the incoherent vector meson production which is sensitive to fluctuations in the transverse density profile of the target. In this study, we have considered two models for the profile functions, with one them considering that the nucleons can have subnucleonic degrees of freedom, denoted hot - spots. Our results demonstrate that the impact of the hot - spots is larger for larger virtualities and lighter nuclei. In particular, future analysis of the ratio between the incoherent and coherent cross sections and the momentum transfer distributions can be useful to constrain the description of the hadronic structure of the nucleus.

%
%

\section{Acknowledgement}
The authors are deeply grateful to the very useful discussions with J. G. Contreras and J. D. Tapia Takaki. VPG is grateful to the members of the Department of Physics and Astronomy of the University of Kansas by the warm hospitality during the initial phase of this study.
JC has been supported by the grant 17-04505S of the Czech Science Foundation (GACR). VPG  was  partially financed by the Brazilian funding
agencies CNPq,  FAPERGS and INCT-FNA (process number
464898/2014-5). MK was supported by the Conicyt Fondecyt grant Postdoctorado
N.3180085 (Chile) and by the grant LTC17038 of the Ministry of Education, Youth and
Sports of the Czech Republic.
Access to computing and storage facilities of the National Grid Infrastructure MetaCentrum provided under the programme CESNET LM2015042 of
the Czech Republic is greatly appreciated.

%
%
\bibliographystyle{utphys}


\end{document}